\documentstyle[12pt]{article}

\input{epsf}
\newcommand{\bc}{\begin{center}}
\newcommand{\ec}{\end{center}}
\newcommand{\bt}{\begin{tabular}}
\newcommand{\et}{\end{tabular}} 
\newcommand{\bdes}{\begin{description}}
\newcommand{\edes}{\end{description}}

\newcommand{\be}{\begin{equation}}
\newcommand{\ee}{\end{equation}}
\newcommand{\bea}{\begin{eqnarray}}
\newcommand{\eea}{\end{eqnarray}}
\newcommand{\non}{\nonumber}

\newcommand{\half}{\frac{1}{2}}
\newcommand{\ba}{\begin{array}}
\newcommand{\ea}{\end{array}}

\newcommand{\bkap}{\mbox{\boldmath $ \kappa $}}

\newcommand{\bvhi}{\mbox{\boldmath $ \varphi $}}

\newcommand{\bbb} { {\bf b}}
\newcommand{\bff} { {\bf f}}
\newcommand{\bu} { {\bf u}}

\newcommand{\bC} { {\bf C}}

\newcommand{\bB} { {\bf B}}
\newcommand{\bF} { {\bf F}}
\newcommand{\bK} { {\bf K}}

\newcommand{\bP} { {\bf P}}

\newcommand{\br} { {\bf r}}
\newcommand{\bR} { {\bf R}}

\newcommand{\bU} { {\bf U}}
\newcommand{\bI} { {\bf I}}

\newcommand{\bx} { {\bf x}}
\newcommand{\bX} { {\bf X}}

\newcommand{\dou}{\partial}

\newcommand{\eqn}[1] {eq.~(\ref{#1})}
\newcommand{\fig}[1] {fig.~\ref{#1}}
\newcommand{\Fig}[1] {Fig.~\ref{#1}}
\newcommand{\tab}[1] {table~\ref{#1}}
\newcommand{\Tab}[1] {Table~\ref{#1}}

\newcommand{\nalpha}{n_\alpha}

\newcommand{\ealpha}{E_\alpha}
\newcommand{\edens}{{\cal E}}
\newcommand{\rcut}{r_{\rm cut}}

\newcommand{\figdir}[1]{./#1 }

\newcommand{\wheretoput}{}
\newcommand{\figcaption}[1]{\caption[#1]{}}

\title{ \Large \sc An Adaptive
Finite Element Approach  to  Atomic-Scale
Mechanics -   The Quasicontinuum Method } 
\author{{\normalsize V. B. Shenoy$^1$, R. Miller$^1$,
E. B. Tadmor$^2$, D. Rodney$^1$, R. Phillips$^1$ and M. Ortiz$^3$ } \\
{\small $^1$Division of Engineering, Brown University, Providence, RI
02912} \\ 
{\small $^2$Division of Applied Science, Harvard University,
Cambridge, MA 02138} \\ 
{\small $^3$Department of Aeronautics, California Institute of Technology, Pasadena, CA 91125}} 
\date{}

\begin{document}
\baselineskip=24pt
\begin{titlepage}
\begin{figure*}
\centerline{\epsfysize=1.0in \epsfbox{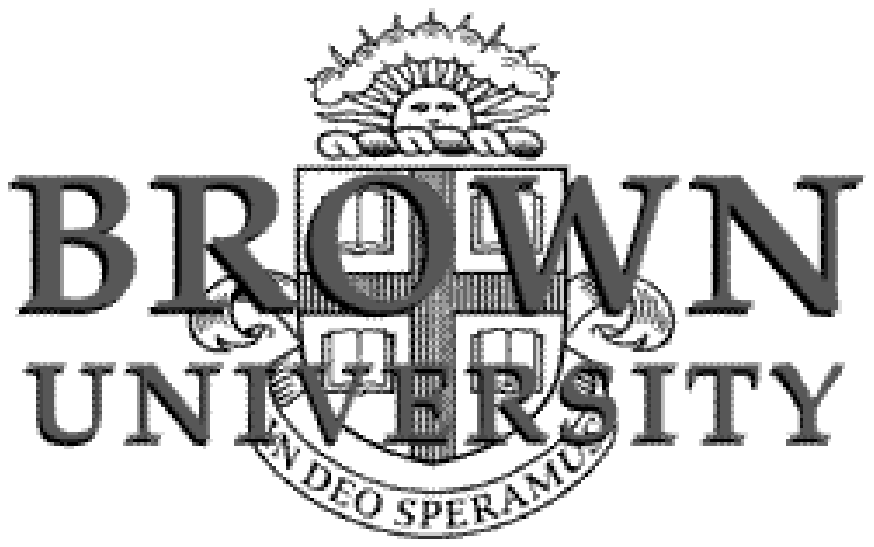}}
\end{figure*}
\Large
\bc
{\bf 
An Adaptive
Finite Element Approach  to  Atomic-Scale
Mechanics -   The Quasicontinuum Method }
\ec
\large
\vspace{2.5in}
\bc
 V.~B.~Shenoy, R.~Miller, E.~B.~Tadmor, D.~Rodney, R.~Phillips, M.~Ortiz 
\ec
\vspace{3.0in}
\bc
Atomistics Group \\
Division of Engineering \\
Brown University \\
Providence, RI 02912
\ec
\end{titlepage}

\maketitle

{\small

}

\begin{abstract}

Mixed atomistic and continuum methods offer the possibility of
carrying out simulations of material properties at both larger length
scales and longer times than direct atomistic calculations.  The
quasi-continuum method links atomistic and continuum models through the
device of the finite element method which permits a reduction of the
full set of atomistic degrees of freedom.  The present paper gives a
full description of the quasicontinuum method, with special reference
to the ways in which the method may be used to model crystals with
more than a single grain.  The formulation is validated in terms of a
series of calculations on grain boundary structure and energetics.
The method is then illustrated in terms of the motion of a stepped
twin boundary where a critical stress for the boundary motion is calculated  
and nanoindentation into a solid containing a subsurface grain boundary to 
study the interaction of dislocations with grain boundaries.
\end{abstract}

\section{Introduction}

A longstanding ambition in the modeling of materials has been that of
rationalizing and predicting the observed mechanical properties of
materials on the basis of an understanding of their constituent
defects.  The advent of ever more  powerful computers alternatives has ushered in
the possibility of carrying out rudimentary calculations of this type
directly on the basis of full atomistic simulations.  However, an
alternative class of models has sought to exploit atomistic insights
without abandoning altogether the powerful resources that are
associated with continuum theories.  One such approach is the
quasicontinuum method (Tadmor, Ortiz and Phillips, 1996) which links
the kinematic constraints, and attendant degree of freedom reduction
offered by the finite element method, to total energies predicated
entirely on atomistic analysis.

The push to develop models of defect interactions has come from
experimental observations on ever-smaller length scales. Recent
micromechanical observations routinely explore problems in which
small numbers of defects are responsible for the mechanical properties
of interest (see, for example, Gerberich, Nelson, Lilleodden, Anderson and Wyrobek, 1996). The experimental advent of nanomechanics as
ushered in by a host of high resolution microscopies such as
high-resolution TEM and atomic scale resolution surface probes such as
the STM and the AFM has led to theoretical demands as well.  One of
the theoretical responses to this challenge has been the attempt to
build simulation tools that allow for the analysis of multiple length
scales simultaneously.  Before turning to the quasicontinuum method
itself, we mention two examples that are antecedents to the approach
advocated here.

The broad class of models known as cohesive zone models have as their
aim the incorporation of constitutive nonlinearity to account either
for the ``core'' material at a crack tip (Barenblatt 1962) or at a
dislocation core (Peierls 1940).  One of the significant outcomes of
these calculations that is especially noteworthy for the present
purpose is the fact that the incorporation of the constitutive
nonlinearity  introduced by the cohesive zone eliminates (or at
least ameliorates) the singularities that are inherited from an
analysis predicated purely on the basis of linear elasticity.  A
recent advance in cohesive zone technology has been the exploitation
of cohesive zone elastic potentials calculated using density
functional calculations (Xu, Argon and Ortiz, 1995), in the study of
dislocation nucleation near a crack tip. As will be more evident below, the
quasicontinuum method takes its cue from the cohesive zone approach in
that it is noted that constitutive nonlinearity, especially as
dictated by an underlying atomistic analysis, that gives this approach
its power.

An alternative class of models that is built around the same insights
as those associated with cohesive zone approaches are those that
effect a linkage between two different spatial regions, one of which
explicitly treats each and every atomic degree of freedom with its
requisite nonlinearity and a second of which is treated either by
traditional linear elasticity or its discrete analog (Kohlhoff,
Gumbsch and Fischmeister (1991), Thomson, Zhou, Carlsson and Tewary
(1992)).  Such methods necessitate the management of boundary
conditions that permit the matching of the two regions.  However, the
logic that stands behind such methods is nearly identical to that advocated
here.  The assertion is that in the immediate vicinity of the defect,
it is essential to have a sufficient level of resolution as to capture
the sometimes complex local atomic rearrangements that characterize
them.  On the other hand, the contention is that far away from the
defect, there is no reason to doubt the efficacy of continuum
mechanics.  In their present incarnations, these mixed methods have
the difficulty in that each problem must be treated on a case by case
basis with the boundary between the fully atomistic region and the
rest of space designed {\it a priori}.  The quasicontinuum method to
be described below is founded on the basis of adaptive strategies in
which, in the course of the calculation, the region of full atomistic
resolution can vary in response to the evolution of both loading and
defects.

The present paper is written with two  clear objectives.
First, it is our aim to generalize the earlier statements of the
quasicontinuum method so as to allow for the treatment of problems
involving more than one grain.  It is well known
that a range of phenomenology in the mechanics of materials including
Hall-Petch type phenomena, grain boundary sliding and Nabarro-Herring
creep,  trace their existence to the
presence of grain boundaries. The formulation as described in an
earlier paper (Tadmor, Ortiz and Phillips, 1996) was noncommittal with
respect to the question of how one might incorporate grain boundaries
as a synthetic element of the material's microstructure and it is a key
aim of the present work to remove that ambiguity.

The second objective of the present paper is to provide the logical
foundation for the generalization of the method to a number of
different contexts.  Work to further extend the quasicontinuum method
is presently in progress in a number of different areas, all of which
rest upon the foundations laid here.  One of the key extensions is to
make the method fully three dimensional, which allows for the
realistic simulation of phenomena such as dislocation junction
formation and dislocation nucleation.  An additional area of
development concerns the need to explicitly evaluate the dynamical
trajectories undertaken in a given process at finite temperature.  In
particular, we see the quasicontinuum method as a possible meeting
point for molecular dynamics and continuum thermodynamics.  Finally,
to allow for the treatment of alloys and even elemental materials such
as silicon, the method has required generalization to situations in
which the crystallography is characterized by the presence of more
than one atom per unit cell.  Although these topics will not be
described in the present paper explicitly, preliminary efforts have
been forged in all of these directions, all of which have been built
around the ideas that are set forth here.

The organization of the remainder of the paper is as follows.  Section
2 presents an overview of the
quasicontinuum method. The detailed
 description of the individual
components of the methodology may be found in Section 3. This will be
followed in section 4 by a discussion of validation of the method,
with full atomistic calculations serving as the benchmark for
success with particular  reference to the use of
quasicontinuum calculations for the study of grain boundary structure
and energetics.  Sections 5 and 6  complete the description with applications
that exploit the capacity of the method to examine deformation problems involving
more than a single grain.

\section{Overview}
In this section, we undertake a description of the quasicontinuum
method, with special attention drawn to the subtleties that attended
the generalization to problems involving more than one grain.  As
discussed in the introduction, one of the fundamental precepts around
which the quasicontinuum method was built was the idea that direct
atomistic simulation has both strengths and weaknesses.  The clear
advantage of the atomistic perspective is that such calculations are
capable of providing the requisite resolution to account for the
highly nonlinear and sometimes counterintuitive atomic arrangements
that are found in the defect core. Indeed, it has been found that in
some circumstances such atomic level details are the source of known
macroscopic anomalies in the material's behavior (Christian 1983).  On
the other hand, the weakness of the atomistic approach is the huge
number of redundant degrees of freedom away from such defects.  The
quasicontinuum method attempts to incorporate both of these insights
by allowing for full atomic scale resolution near defect cores while
exploiting a coarser description further away.

We begin with an overview of the conceptual elements of the
quasicontinuum method as described in Tadmor, Ortiz and Phillips
(1996).  The key idea is that of kinematic slavery in which by virtue
of the finite element method (FEM), the positions of the majority of atoms
are entirely constrained and determined only by the displacements of
the nodes tied to the element of which they are a part.  Once the
geometric disposition of the body is established, the problem becomes
one of determining the total energy.  From a traditional continuum
mechanics viewpoint, such as that offered by the linear theory of
elasticity, the total energy may be computed on the basis of a
phenomenological constitutive law such as Hooke's law.  The aim in Tadmor,
Ortiz, Phillips (1996), by way of contrast, has been to use atomistic
calculations to inform the energetic statement of the continuum
mechanics variational principle.  This step offers the 
constitutive nonlinearity alluded to above, and allows naturally for
the emergence of lattice defects such as dislocations.

We now give a formal description of the formulation of the
quasicontinuum method generalized to account for the presence of
multiple grains, restricting attention to those problems in which the
undeformed state of the body is polycrystalline.  We take the view that
the body whose disposition is of interest should be thought of as a
collection of a possibly huge number, $N$, of atoms. In  
\fig{pdef}, we show the body which is imagined to be built up of a variety of
different grains with Bravais lattice vectors  schematically
indicated. The presence of a crystalline reference configuration is
exploited in the sense that for many regions of the crystal, it is
unnecessary to save lists of atomic positions since they can be
generated as needed by exploiting the crystalline reference state.  A
given atom in the reference configuration is specified by a triplet of
integers ${\bf l}=(l_1,l_2,l_3)$, and the grain to which it belongs.
The position of the atom in the reference configuration is then given as, 
\bea 
\bX(l)=\sum_{a=1}^3 l_a \bB_a^\mu+\bR^\mu,
\label{brav}
\eea
where $\bB_a^\mu$ is the $a^{\mit th}$ Bravais lattice vector
associated with grain $G_\mu$ and $\bR^\mu$ is the position of a reference
atom in grain $G_\mu$ which serves as the origin for the atoms in 
grain $G_\mu$. 

Once the atomic positions have been
given, from the standpoint of a strictly atomistic perspective, 
the total energy is given by the  function 
\bea
E_{tot}=E_{exact}(\bx_1, \bx_2, \bx_3, ..., \bx_N)=
E_{exact}(\{\bx_i\}),  \label{ExactEnergy}
\eea
where $\bx_i$ is the deformed position of atom $i$. We adopt the convention
that capital letters refer to the undeformed configuration while lower
case letters refer to the deformed configuration.
The energy function in \eqn{ExactEnergy} depends explicitly upon each and 
every microscopic degree of freedom, and as it stands
becomes intractable once the number of atoms exceeds one's current
computational capacity.
The problem of determining the minimum of the potential energy is in the
context noted above  nothing more than a statement of conventional lattice statics.
We will now proceed to construct the formulation of the method in a step-by-step
fashion.  We begin with the introduction of kinematic constraints which
have the effect of reducing the number of degrees of freedom
being accounted for by selecting a small subset of $R$ atoms from the
total set of $N$ atoms. These atoms serve as ``representative atoms'', 
and remain the only unconstrained degrees of freedom in the problem.
A finite element mesh with nodes corresponding to the positions of the
representative atoms is then defined. By virtue of finite element 
interpolation we may compute the total energy from the equation
given above, but now with a substantial fraction of the atoms participating
geometrically as nothing more than drones since their positions are entirely 
determined by the displacements of the adjacent  nodes.
Quantitatively, if $\alpha$ is an index over the representative atoms,
then the interpolated position ${\bf x}_i^{int}$ of any other atom
$i$ may be obtained by
\bea
{\bf x}_i^{int} = \sum_\alpha N_\alpha(\bX_i) \bx_\alpha,
\eea 
where $ N_\alpha(\bX_i) $ is the finite element shape function
centered around the representative atom $\alpha$ (which is also a FEM node)
 evaluated at the undeformed position $\bX_i$ of atom $i$. 
In particular, we may write the total energy as
\bea
E_{tot}=E_{exact}(\bx_1^{int}, \bx_2^{int}, ...,\bx_N^{int})=
E_{exact}(\{\bx_i^{int}\}). \label{ExactConstraint}
\eea

At this stage not much has been gained since the computation of the
total energy is still predicated upon a knowledge of all of the atomic
positions, though now many of the atomic positions are constrained.
We now make the additional assumption that the energy may be written
in a form that is additively decomposed, such that 
\bea
E_{tot}=\sum_{i=1}^N E_i , \label{EnergyDecompose}
\eea
which
presupposes the existence of  well-defined site energies $E_i$, and
is  typical in  many  current atomistic formulations such as the
embedded-atom method (EAM).  The summation runs over all atoms in the
solid.  Because of this, the sum written above remains intractable in
the sense that if we interest ourselves in computing the total energy,
we are still obliged to visit each and every atom.  The spirit of the
problem that we are faced with is now identical to that of numerical
quadrature, and what we require at this point is a scheme for
approximating the sum given above by summing only over the
representative atoms with appropriate weights selected so as to
account for differences in element size and environment. In
particular, we desire 
\bea 
E_{tot} \approx \sum_{\alpha=1}^R n_\alpha E_\alpha. \label{EnergyApprox} 
\eea 
The crucial idea embodied in this
equation surrounds the selection of some set of representative atoms,
each of which are intended to characterize the energetics of some
spatial neighborhood within the body as indicated by the weight
$n_\alpha$.  As yet, the statement of the problem is incomplete in
that we have not yet specified how to determine the summation weights,
$n_\alpha$. We treat the problem of the determination of $n_\alpha$ in a
manner analogous to determination of quadrature weights in the
approximate computation of definite integrals. In the present context
the goal is to approximate a finite sum (``definite integral'' on the
lattice) by an appropriately chosen quadrature rule where the
quadrature points are the sites of the representative atoms. Physically
the quantity $n_\alpha$ may be interpreted as the ``number of atoms
represented'' by the representative atom $\alpha$. The quadrature
rule (\eqn{EnergyApprox}) is designed such that  in the 
limit in which the finite element mesh is refined all the way down to
the atomic scale (a limit we denote as fully refined) each and
every atomistic degree of freedom is accounted for, and the quadrature
weights are unity (each representative atom represents only itself).
On the other hand, in the far field regions where
the fields are slowly varying, the quadrature weights reflect the
volume of space (which is now proportional to the number of atoms)
that is associated with the representative atom. The details of this
procedure may be found in section \ref{redrep}.

\begin{figure}\wheretoput
\centerline{\epsfysize=2.5truein \epsfbox{\figdir{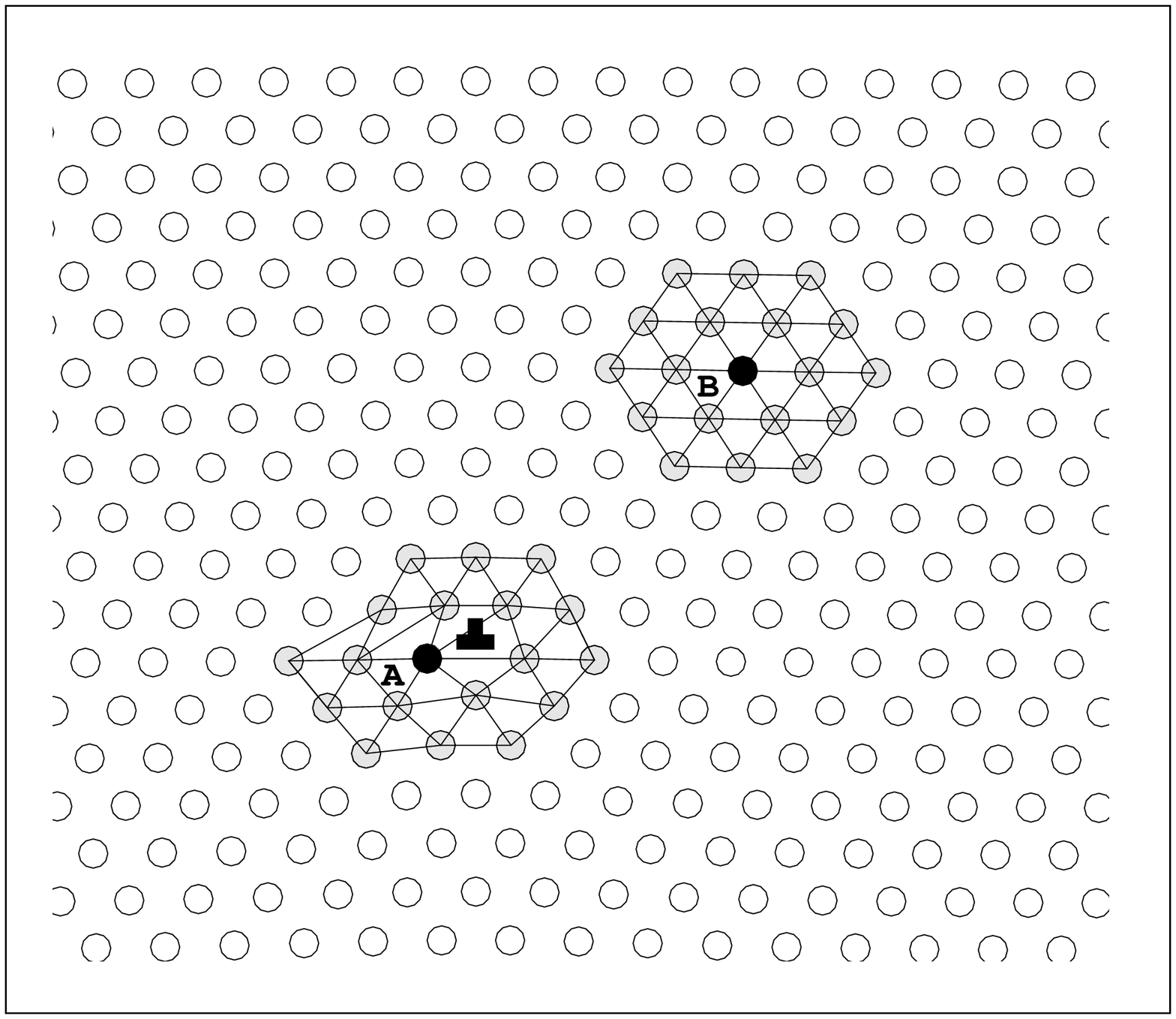}}}
\figcaption{Atomic structure near the core of a Lomer dislocation in
Al. The atom $A$ in the core region experiences an inhomogeneous
environment while the environment of atom $B$ is  nearly homogeneous.}
\label{ratmA}
\end{figure}

The description given  above describes the essence of the formulation
as it is currently practiced.  We now describe an additional energetic
approximation that simplifies the energy calculations and also makes
it possible to formulate boundary conditions which mimic
those expected in an elastic continuum. 
The essential idea is motivated  by \fig{ratmA}, which 
depicts the immediate neighborhood of a dislocation core.
In particular, for this Lomer dislocation
we note the characteristic
geometric signature of the core, namely, the pentagonal group 
of atoms in the core region.  We now focus our attention on the 
environments of two of the atoms in this figure, one (labeled $A$) in the 
immediate core region, and
the other (labeled $B$) in the far field of the defect.
It is evident that the environment of atom $A$ is nonuniform, and that
each of the atoms in that neighborhood experiences a distinctly different 
environment.
By way of contrast, atom $B$ has an environment that may be thought
of as emerging from a uniform deformation, and each of the atoms in
its vicinity sees a nearly identical geometry.

  As a result of these geometric insights, we have found
it convenient to compute the energy $E_\alpha$ from an atomistic
perspective in two different ways, depending upon the nature of the
atomic environment of the representative atom $\alpha$.  Far from the
defect core, we exploit the fact  that the atomic environments are nearly
uniform by making a {\it local} calculation of the energy in which it
is assumed that the state of deformation is homogeneous and can be
well-characterized by the local deformation gradient $\bF$.  To
compute the total energy of such atoms, the Bravais lattice vectors of
the deformed configuration $\bbb_a$ are obtained from those in the
reference configuration $\bB_a$ via $\bbb_a=\bF \bB_a$. Once the
Bravais lattice vectors are specified, this reduces the computation of
the energy to a standard
exercise in the practice of lattice statics.

On the other hand, in regions that suffer a state of deformation that
is nonuniform, such as the core region around atom $A$ in
\fig{ratmA}, the energy is computed by building a crystallite that
reflects the deformed neighborhood from the interpolated displacement
fields.  The atomic positions of each and every atom are given 
exclusively by $\bx=\bX+\bu(\bX)$, where the displacement field $\bu$
is determined from finite element interpolation. This ensures that a
fully nonlocal atomistic calculation is performed in regions of rapidly
varying $\bF$. An automatic criterion for determining whether to use the local or nonlocal rule to
compute a representative atom's energy  based on
the variation of deformation gradient in its vicinity will be
presented in section \ref{energy}.  The distinction between local and
nonlocal environments has the unfortunate side effect of introducing
small spurious forces we refer to as ``ghost'' forces at the interface
between the local and nonlocal regions. A correction for this problem
is also discussed in section \ref{energy}.

Once the total energy has been computed and we have settled both the
kinematic and energetic bookkeeping, we are in a position to determine
the energy minimizing displacement fields. As will be discussed below,
there are a number of technical issues that surround the use of either
conjugate gradient or Newton-Raphson techniques.  Both of these
techniques are predicated upon a knowledge of various derivatives of
the total energy with respect to nodal displacements and are explained
in section \ref{eneminimize}.

As noted in the introduction, one of the design criteria in the
formulation of the method was that of having an adaptive capability
that allowed for the targeting of particular regions for refinement in
response to the emergence of rapidly varying displacement fields.  For
example, when simulating nanoindentation, the indentation process leads
to the nucleation and subsequent propagation of dislocations into the
bulk of the crystal.  To capture the presence of the slip that is tied
to these dislocations, it is necessary that the slip plane be refined
all the way to the atomic scale.  The adaption scheme to be described
in section \ref{adaption}  allows for the natural emergence of such mesh refinement as an
outcome of the deformation history.

The goal of the present section has been to elucidate the key
conceptual elements involved in using the quasicontinuum
method. However, as was noted above, certain features in the
formulation involve subtleties demanding further attention.  In
the following sections, we undertake a more detailed analysis of some of
those issues.

%%%
%%%
\section{Details of the Methodology}

\subsection{Reduced Atomic Representation}
\label{redrep}
\begin{figure}\wheretoput
\centerline{\epsfysize=4.0truein \epsfbox{\figdir{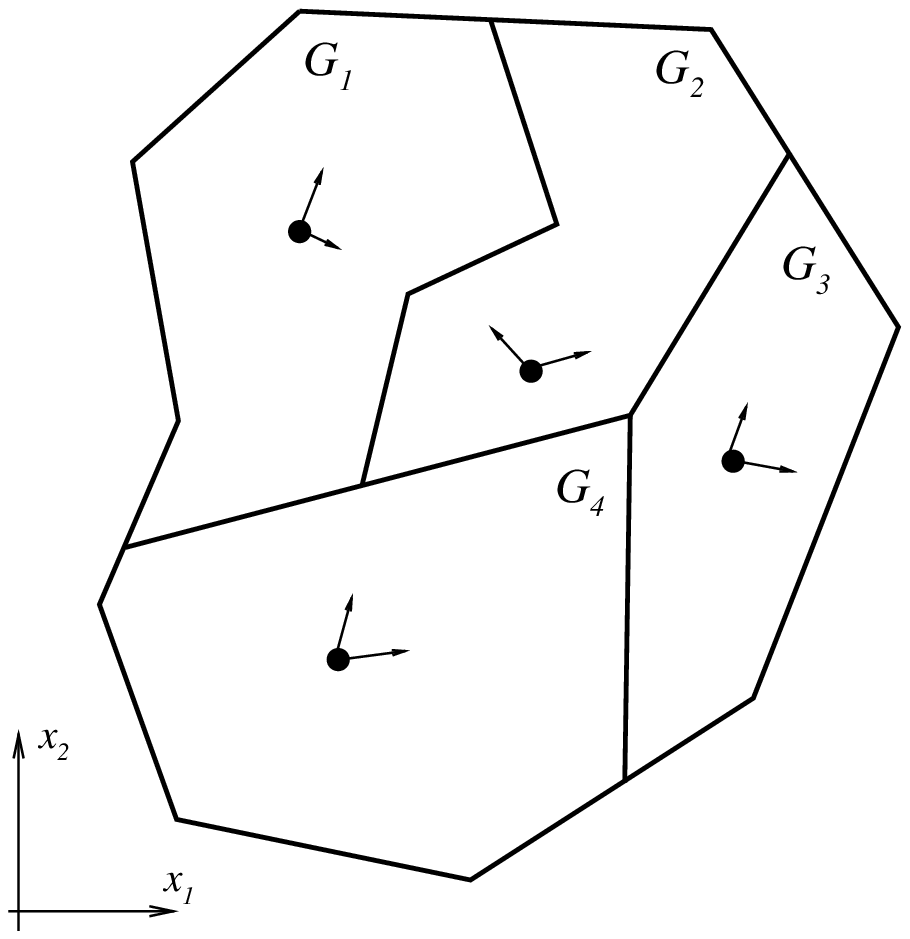}}}
\figcaption{Illustration of a crystalline solid made up of grains $G_\mu$
with a reference atom in each grain and an associated set of Bravais
lattice vectors.}
\label{pdef}
\end{figure}

In this section we discuss the selection of the representative atoms,
and the construction of an expression of the total energy that depends
only on the degrees of freedom of the representative atoms.
We consider the undeformed body to be a {\em crystalline solid}, i.e., a collection of
$N$ atoms, which occupy a region $B_0$ and may be arranged in many grains
$G_\mu$ (see \fig{pdef}). The undeformed position $\bX_i$ of any atom $i$ is obtained
from the coordinate of a reference atom and an associated set of Bravais
lattice vectors that are specified for each grain as discussed earlier
in (\eqn{brav}). The deformed configuration
of the body is described by a displacement function $\bu$ which depends
on $\bX$ and the deformed position of any atom $i$ can be obtained as
\bea
\bx_i = \bX_i + \bu_i,
\eea
where $\bu_i = \bu(\bX_i)$. 

On loading the solid, the equilibrium configuration of the body is
defined by the set of displacements $\bu_i$ which minimizes the
potential energy function
\bea 
\Pi(\bu) = E_{tot}(\bu_1,...,\bu_N) - \sum_{i = 1}^N \bff_i
\cdot \bu_i, 
\label{eexact} 
\eea
where $E_{tot}$ is the total energy of the system obtained from an
atomistic formulation, $\bff_i$ is the external force acting on the
atom $i$ and $\bu$ satisfies the essential boundary conditions of
the problem. This is the well-known method of lattice statics (LS). 
Now, as stated earlier, we assume that $E_{tot}$ can be decomposed as 
a sum over the energies of individual atoms $E_i$, i.e.,
\bea
E_{tot}(\bu_1,...,\bu_N) = \sum_{i = 1}^N  E_i(\bu_1,...,\bu_N),
\eea
and \eqn{eexact} becomes
\bea
\Pi(\bu)  =   \sum_{i = 1}^N  E_i(\bu_1,...,\bu_N) - \sum_{i = 1}^N \bff_i \cdot \bu_i.  \label{evp}
\eea
 Many of the conventional atomistic formulations (such as the
embedded-atom method) admit such a
decomposition, although it is not
admissible in the more sophisticated density functional approach.  
For example, the embedded-atom method (Daw, Baskes 1986) provides that for a homonuclear material the energy at site $i$ is given by
\bea
E_i = \half\sum_{j} \phi(r_{ij}) + f(\rho_i),
\eea
where $r_{ij}$ is the distance from atom $i$ to the neighbor $j$, 
$\phi$ is a pairwise interaction, $\rho_i$ is the electron density at the
site $i$ and $f(\rho)$ is the  embedding energy. The potentials are assumed
to have a cutoff radius of $\rcut$, i.e., any atom interacts directly only with those atoms within a distance $\rcut$ from it.

\begin{figure}\wheretoput
\centerline{\epsfysize=2.5truein \epsfbox{\figdir{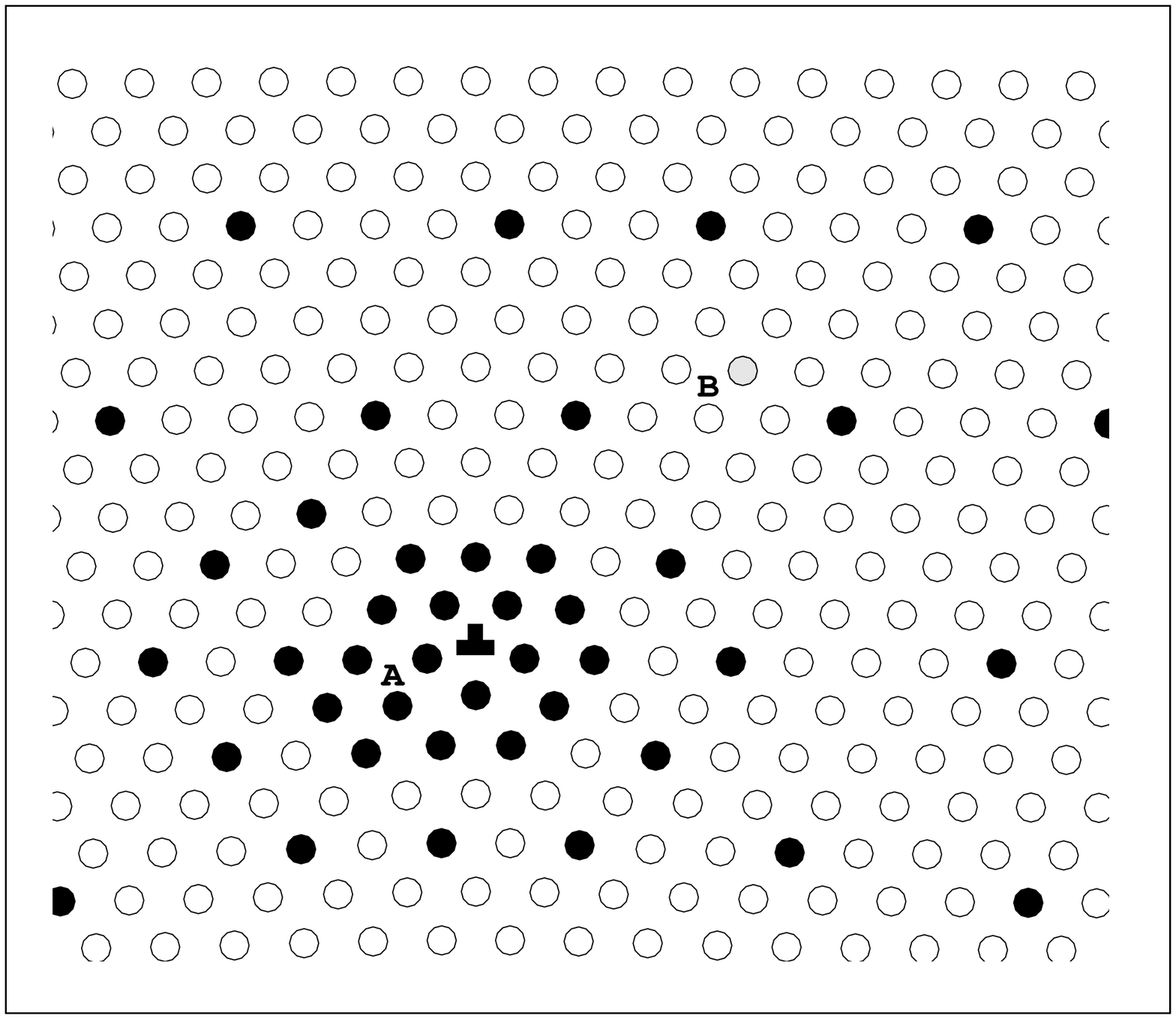}} \epsfysize=2.5truein \epsfbox{\figdir{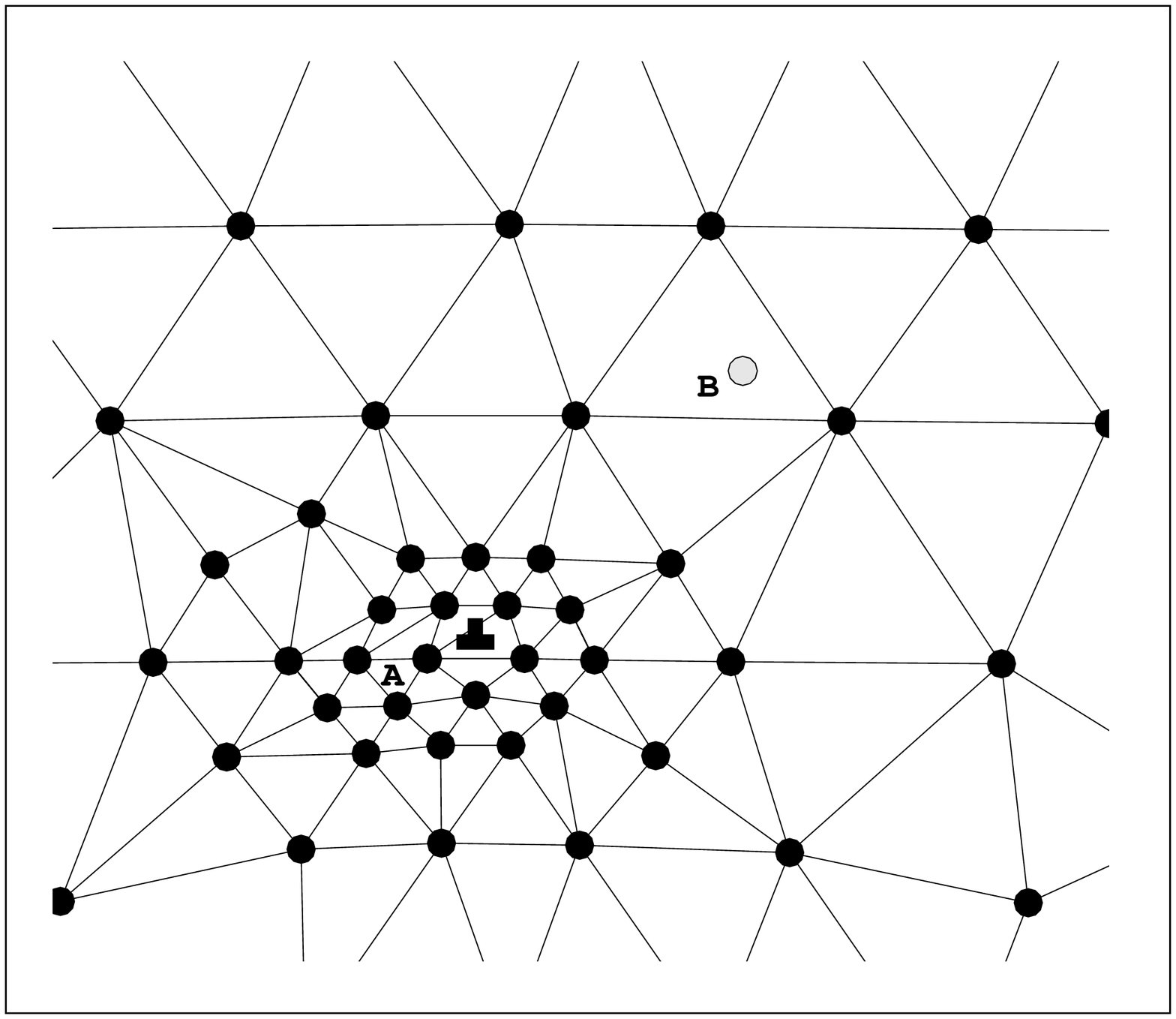}}
}
\centerline{ (a)  \hspace{2.80truein} (b)}
\figcaption{(a) Selection of representative atoms in the case of the Lomer
dislocation in Al. Filled circles are representative atoms while open circles correspond to atoms whose positions are constrained. (b) Finite element mesh constructed to simulate the Lomer
dislocation in Al for the choice of
representative atoms shown in (a). }
\label{ratmB}
\end{figure}

The variational principle associated with \eqn{evp} provides the solution $\bu_i$
and may lead to nonlinear minimization problems with intractable
numbers of degrees of freedom. This motivates the need to formulate
approximation strategies that preserve the essential details of the
problem while requiring fewer degrees of freedom. The first step in
our approximation method is the selection of a subset of $R$ atoms
($R\ll N$) called {\em representative atoms}, to describe the
kinematics {\em and} energetics of the body.  To motivate the reasoning
underlying this approach, we revisit \fig{ratmA} discussed earlier
where the atomic structure in the vicinity of a relaxed  Lomer
dislocation in fcc aluminum is presented.  Two atoms have been
highlighted along with their respective environments.  Atom $A$ lies
at the dislocation core itself, while atom $B$ is further away. In the
region around atom $A$, the deformation fields are changing rapidly on
the scale of atomic distances. The non-uniform nature of the
deformation near the dislocation core implies that all atoms in that
region experience completely different environments.  At the
same time, many of the atoms in the vicinity of atom $B$ experience
environments nearly identical to that of atom $B$, and thus are nearly
energetically equivalent to atom $B$. This conclusion naturally leads
to the concept of representative atoms mentioned above.  In this case,
atom $B$ could well represent several of its neighbors, due to their
similar environments, while all atoms near $A$ have to be chosen as
representative atoms.  Where the deformation gradients are large and
quickly varying on the atomic scale more representative atoms will be
selected, while further away fewer will be explicitly considered. Such
a representation is presented in \fig{ratmB}a for the same dislocation
core structure. We see that near the core region all atoms are
represented, while further away, where the deformation is more
homogeneous, the density of representative atoms is reduced.

The displacements \{$\bu_\alpha$\} of the representative atoms are  the relevant degrees of freedom of the system. The next task is to
construct an approximate energy function $\Pi_h$ that depends {\em
only on \{$\bu_\alpha$\}}.  To achieve this we first need a kinematic
description of the body, i.e., we need a tool to describe the deformed
positions of every atom in the body when the displacements of the
representative atoms are known. This is achieved by the construction
of a finite element mesh ($\Omega_e, e = 1...M$, where $M$ is the
number of elements) with the representative atoms as nodes
(cf. \fig{ratmB}b).  The deformed position of any atom in the model is
then obtained by interpolation using the finite element shape
functions and the displacements of the representative atoms; the
deformed configuration of the body may thus be completely described.
We have chosen to use linear triangular finite elements (i.e., the
deformation gradient is constant in each element), which are generated
using the constrained Delaunay triangulation (Sloan 1993). This
triangulation allows for the convenient treatment of non-convex, multiply
connected regions.

The energy of the body is  described by \eqn{EnergyApprox}, and
requires a knowledge of $n_\alpha$.  As explained above, these
quantities may be thought of as quadrature weights for the
summation  on the discrete lattice. We now discuss
the computation of $n_\alpha$. Let $g$ be a real
valued function on the lattice with  $g(\bX_i)$ being its value at the
site $i$. We define 
\bea 
{\cal S}(g) = \sum_{i=1}^{N} g(\bX_i),
\eea 
as the sum of $g$. If this sum is to be evaluated
using the value of $g$
only at the representative atoms, $g(\bX_\alpha)$, and a quadrature rule,
we have 
\bea {\cal S}_h(g) = \sum_{\alpha=1}^{R} \nalpha g(\bX_\alpha).  
\eea 
The
values of $\nalpha$ may now be determined by insisting that 
\bea {\cal
S}(g_\beta) = {\cal S}_h(g_\beta) \label{QuadCriterion}
\eea 
for some functions
$g_\beta \;\; (\beta=1...R)$ chosen {\it a priori}, i.e., by
insisting that the quadrature rule should sum the functions
$g_\beta$ exactly. There are many possible choices of $g_\beta$, some of
which we discuss below.
\begin{enumerate}
\item{{\it Voronoi characteristic functions}: The function $g_\beta$ is chosen to be the characteristic function of the Voronoi cell $V_\beta$  (Okabe, Boots and Sugihara, 1992), associated with representative atom $\beta$,
\bea
\chi^V_\beta (\bX_i) & =&  1, \;\;  \forall \; \bX_i \in V_\beta \non \\
                  & = & 0, \;\; \mbox{otherwise}.        
\eea
Setting $g_\beta = \chi^V_\beta$ and applying the relation in  
\eqn{QuadCriterion} we obtain $n_\beta$ to be
\bea
n_\beta = {\cal S}_h(\chi^V_\beta) =   {\cal S}(\chi^V_\beta)) = \sum_i \chi^V_\beta(\bX_i).
\eea
It is easily seen that $\nalpha$ admits a simple interpretation as the
total number of atoms in the Voronoi cell of representative atom $\alpha$.
 }

\item{ {\it Patch characteristic functions}: We define a patch
($P_\beta$) surrounding a representative atom/node $\beta$ as
the polygon constructed by joining the centroids and midpoints of the
edges of the elements incident on the representative
atom.  The function
$g_\beta$ is chosen to be the characteristic function of the patch
$P_\beta$ and it is now immediate that
\bea 
n_\beta =  {\cal S}(\chi^P_\beta) = \sum_{i=1}^{N} \chi_\beta^P(\bX_i)
\eea
and again $n_\beta$ may be interpreted as the number of atoms
contained in the patch $P_\beta$.
}

\item{ {\it FE shape functions}: Here we choose the function
$g_\beta$ to be the finite element shape function $N_\beta$, and
in a manner similar to the above two cases it follows that 
\bea
n_\beta = {\cal S}(N_\beta) = \sum_i N_\beta(\bX_i).
\eea
}

\end{enumerate}
Some remarks are in order. First, note that in each case, the
functions $g_\beta$ are chosen so that their value is unity at the
representative atom $\beta$ and vanishes at all other representative
atoms. Second, the first two alternatives require the explicit
construction of a tessellation (either Voronoi cells or the
patches). Third, all the alternatives provide the quadrature weight
to be unity at a representative atom situated in a fully-refined
region. It has  been found that the results of the calculations
are insensitive to the choice of the different schemes.

Armed with the values of $\nalpha$ and \eqn{EnergyApprox} the approximate
energy function  $\Pi_h$ may be written as
\bea
\Pi_h(\bu) = \sum_{\alpha = 1}^{R} \nalpha \ealpha
(\bu_1,...,\bu_R) - \sum_{\alpha=1}^{R} \nalpha \bar{\bff}_\alpha
\cdot \bu_\alpha , \label{avp} 
\eea 
where $ \bar{\bff} $ is the average force on representative atom $\alpha$,
and the subscript $h$ refers to the approximation introduced by the finite
element partitioning. In the event that one chooses all the
atoms in the model the energy  in \eqn{avp} reduces to the
exact function of \eqn{evp} and one recovers lattice statics.

%%%
%%%
\subsection{Computation of Representative Atom Energies}
\label{energy}
The energy of any representative atom may be computed by creating a
list of its neighbors (we call such a list the {\em 
representative crystallite}), and obtaining the deformed positions of these
neighbor atoms using the finite
element interpolation. This approach will require the explicit
neighbor list computation of each of the representative atoms and
proves to be very time consuming. A more efficient strategy can be
formulated as follows. Consider the atom $B$ in \fig{ratmA}. This atom
experiences only a slowly varying  deformation, and
its energy can be well  approximated by that computed using the 
local deformation gradients and the Cauchy-Born rule (Ericksen 1984 and references therein) which states that
the atoms in a deformed crystal will move to positions dictated by the
existing gradients of displacements. 

On the other hand, in the region around atom $A$, the deformation
fields are changing rapidly on the scale of atomic distances. This
observation suggests a division of the representative atoms into two
classes (a) Nonlocal atoms whose energies are computed by an
explicit consideration of all its neighbors and (b) Local atoms
whose energies are computed from the local deformation gradients using
the Cauchy-Born rule.  The former type of representative atoms are
essential to capture the atomistic nature of defect cores and
interfaces while the latter represents the continuum limit and is
used in regions of the solid undergoing only a near homogeneous
deformation. It is emphasized that the local formulation is an
approximation that provides for efficient computation.

We now discuss the criterion that decides if a given representative
atom is to be treated as local or nonlocal. The representative crystallite of a representative atom
experiences various deformation gradients arising from the different elements that
surround it. A state of deformation near a representative atom is
near homogeneous if the deformation gradients that it senses from
the different elements are nearly equal. This can be characterized by
the inequality  
\bea 
\max_{a,b;k} |\lambda^a_k - \lambda^b_k| < \epsilon_L, \label{lcrit} 
\eea 
where $a$ and $b$ run
over all elements that are within some radius $r_s$ of the
representative atom discussed presently, $\lambda^e_k$ is the
$k^{\mbox{th}}$ eigenvalue of the right stretch tensor $\bU_e$ 
(see for example, Chadwick 1976) obtained from the deformation gradient $\bF_e$ in element $e$,
and $\epsilon_L$ is an empirically selected constant.  This criterion
implies that the error in the computation of the energy using the
local deformation gradients when compared with a fully atomistic
calculation of the energy is within a specified tolerance. Thus all atoms
for which \eqn{lcrit} is satisfied are treated as local atoms while
for the remaining atoms the nonlocal rule is used to compute the
energy.  In addition, any atom which is within $r_s$ of a surface or
interface of interest is made nonlocal.

Using the above criterion, the total number of nonlocal atoms in the 
model is determined by two prescribed parameters, the tolerance on 
the eigenvalues, $\epsilon_L$, and the range of nonlocal influence, $r_s$.
For correct surface and interfacial energy, atoms within $\rcut$ of the
interface must be nonlocal (i.e. $r_s=\rcut$). For correct
forces and stiffness (promising a correct relaxed configuration) each of
these atoms must be embedded in a further $\rcut$ radius of nonlocal
atoms, bringing the total necessary nonlocal padding to $r_s=2\rcut$.
A lesser padding will result in less demanding computational models with
fewer degrees of freedom at the expense of more approximate interfacial 
and defect core structures.

The total energy (\eqn{avp}) can now be separated into its local and
nonlocal contributions,
\bea
\Pi_h(\bu) = 
\sum_{\alpha = 1}^{R_L} \nalpha E_\alpha^{\mit loc} (\bF_1,...,\bF_M) + 
\sum_{\beta = 1}^{\mit R_{NL}} n_\beta E_\beta (\bu_1,...,\bu_R) - 
\sum_{\alpha=1}^{R} n_\alpha \bar{\bff}_\alpha
\cdot \bu_\alpha , \label{avp_split} 
\eea 
where there are $R_L$ local atoms and $R_{NL}$ nonlocal atoms (such that
$R_L+ R_{NL}=R$).

The procedure used in the calculation of energies of the local atoms
from the local gradients of deformation is taken up presently.
Consider representative atom $\alpha$ experiencing near homogeneous
deformation.  If $\nalpha^e$ denotes ${\cal S}(\chi(\Omega_e) \cdot \psi_\alpha)$, which may be interpreted as the number of atoms  associated with the element $\Omega_e$
represented by the representative atom $\alpha$, then the energy $E_\alpha^{\mit loc}$ 
is approximated by
\bea
\nalpha E_\alpha^{\mit loc} = \sum_{e = 1}^{M} \nalpha^e {\edens}(\bF_e), 
\;\;\;\;  
\nalpha = \sum_{e=1}^{M} \nalpha^e, \label{local}
\eea
where $\edens(\bF)$ is the energy of a single atom experiencing a
homogeneous deformation given by the deformation gradient tensor
$\bF$. 

The nonlocal energy $E_\beta(\bu_1,...,\bu_R)$ is computed as it would be
in a standard atomistic calculation. The energy of atom $\beta$ is a function
of the positions of all atoms falling inside its cutoff sphere for the given deformation.
We assume there are $m_\beta$ such neighbors which occupy positions
$\bX^j_\beta, j=1,2,...,m_\beta$ in the undeformed configuration. The
deformed positions of these atoms relative to atom $\beta$ will be,
\bea
\br^j_\beta=\bX^j_\beta+\bu^j_\beta-\bX_\beta-\bu_\beta,
\label{rjb}
\eea
where $\bX_\beta$ and $\bu_\beta$ are the coordinates and displacements at
atom $\beta$ and $\bu_j^\beta$ is the displacement interpolated to the site
of neighbor $j$ of atom $\beta$ and is given by,
\bea
\bu^j_\beta=\sum_{\alpha=1}^R\bu_\alpha N_\alpha(\bX^j_\beta),
\label{ujb}
\eea
where $\bu_\alpha$ is the displacement at representative atom/node $\alpha$, $N_\alpha$ is the 
associated finite element interpolation function.
Substituting \eqn{local} into \eqn{avp_split} and specifically accounting
for the dependence of $E_\beta$ on $\br^j_\beta$ we have,
\bea
\Pi_h(\bu) = 
\sum_{\alpha = 1}^{R_L} \sum_{e=1}^M \nalpha^e {\edens}(\bF_e) + 
\sum_{\beta = 1}^{\mit R_{NL}} n_\beta E_\beta (\br^1_\beta,...,\br^{m_\beta}_\beta) - 
\sum_{\alpha=1}^{R} n_\alpha \bar{\bff}_\alpha
\cdot \bu_\alpha. \label{avp_split1} 
\eea 
The sums in the first expression of \eqn{avp_split1} can be reversed so
that,
\bea
\Pi_h(\bu) = 
\sum_{e=1}^M \nu_e {\edens}(\bF_e) + 
\sum_{\beta = 1}^{\mit R_{NL}} n_\beta E_\beta (\br^1_\beta,...,\br^{m_\beta}_\beta) - 
\sum_{\alpha=1}^{R} n_\alpha \bar{\bff}_\alpha
\cdot \bu_\alpha, \label{avp_split2_tmp} 
\eea 
where 
\bea
\nu_e=\sum_{\alpha = 1}^{R_L} n_\alpha^e
\eea 
is the total number of atoms (associated with local representative atoms) falling in element $e$.

Although \eqn{avp_split2_tmp} is a complete description of the
approximate energy function, and represents a mathematically
consistent formulation, it leads to noisy solutions in the
presence of nonlocal representative atoms with weights exceeding
unity. Expressed differently, solutions are smoother when nonlocal
representative atoms are present only in the fully refined
regions. The cause of this noise or error may be traced to the
non-uniformity in the constrained kinematics due to a non uniform mesh
(this effect has been noted by Tadmor, Ortiz and Phillips (1996) and
detailed in Tadmor (1996)). The remedy for this
problem is to fully refine the mesh around atoms that are treated
nonlocally, and as a result the weights associated with the nonlocal
atoms will be unity. The resulting approximate energy function
\eqn{avp_split2_tmp} reduces to 
\bea
\Pi_h(\bu) = 
\sum_{e=1}^M \nu_e {\edens}(\bF_e) + 
\sum_{\beta = 1}^{\mit R_{NL}}  E_\beta (\br^1_\beta,...,\br^{m_\beta}_\beta) - 
\sum_{\alpha=1}^{R} n_\alpha \bar{\bff}_\alpha
\cdot \bu_\alpha, \label{avp_split2} 
\eea

\begin{figure}\wheretoput
\centerline{\epsfysize=3.0truein \epsfbox{\figdir{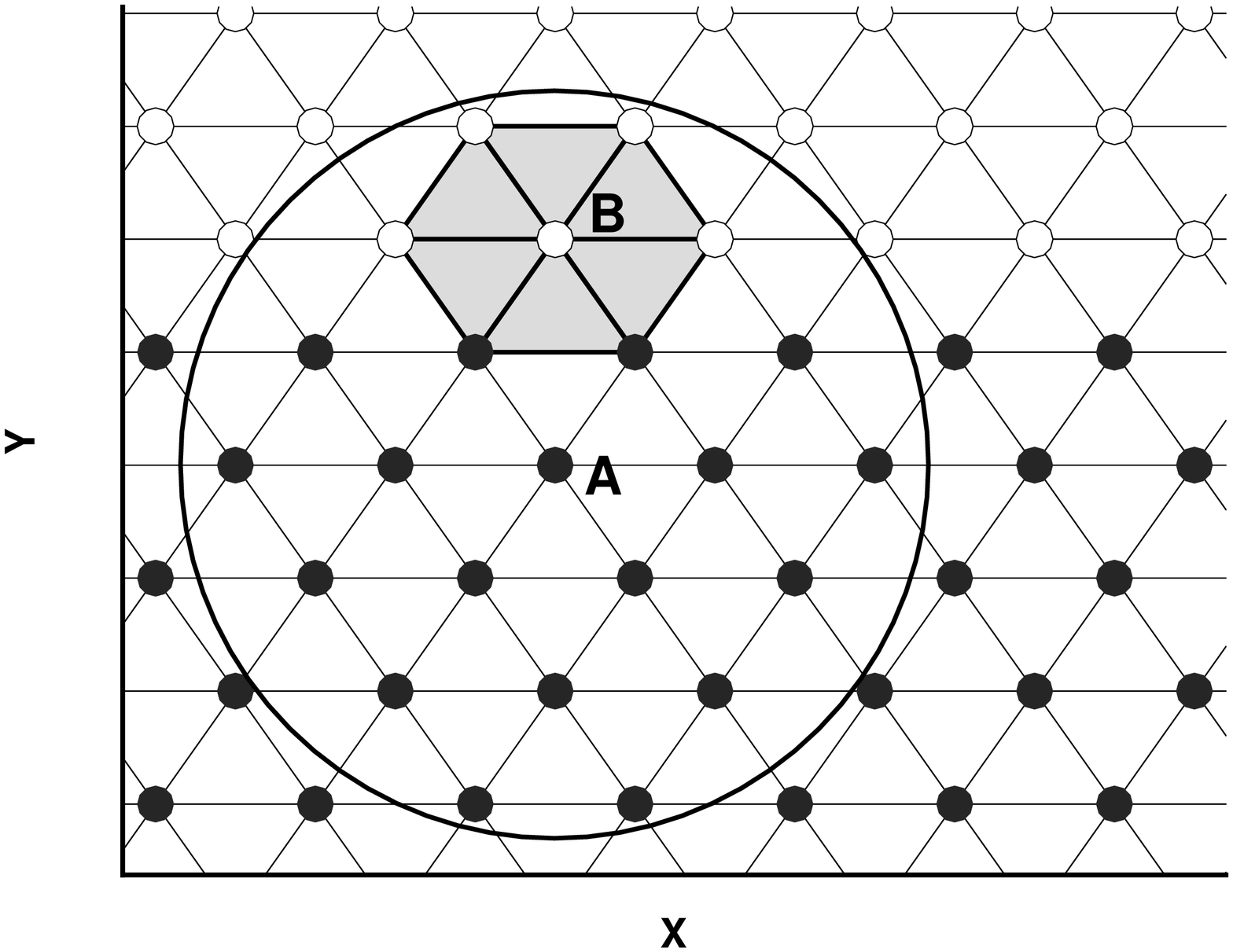}}}
\figcaption{Illustration of  the origin of ``ghost'' forces.}
\label{ghost}
\end{figure}

In the type of configurations we have studied, nonlocal atoms tend to
appear in groups we refer to as clumps, surrounded by local
atoms. The local and nonlocal formulations are not completely
compatible and, although there is no seam in the formulation of the
energy function, non-physical forces arise on atoms in
transition zones between local and nonlocal regions, even when the
crystal is undeformed.  To illustrate this point, consider a fully
refined mesh as shown in \fig{ghost}. The  atoms represented by open circles 
are local atoms
whereas the dark ones are nonlocal. The circle drawn around nonlocal
atom $A$ represents its cutoff sphere. On the one hand, as local atom
$B$ lies inside the sphere, its motion affects $A$ and $\dou E_A^{NL}
/ \dou \bu_B \neq 0$, which corresponds to a force acting on $B$ due
to $A$. On the other hand, the energy of $B$ is computed locally and
depends only on atoms which share a common element with $B$ (the
shaded elements in \fig{ghost}). Therefore, $\dou E_B^L / \dou \bu_A =
0$ and the motion of $A$ does not lead to forces on $B$. These imbalances result in non-physical
forces on $A$ and $B$, which we call ``ghost'' forces. Their order of
magnitude is 0.1 eV/\AA\ and can lead to energy relaxation of
order 0.005 eV. The principal reason for this erroneous unbalanced
forces is the fact that our procedure has focussed on approximating
the energy and not the forces.

In order to avoid the ghost forces, we have to correct the
forces acting on atoms in the transition zones. This is achieved by demanding that  forces
acting on any atom be computed using only the formulation which
corresponds to its status, as if the atom was in a fully local or
nonlocal region. For example, the nonlocal term $\dou E_A^{NL} / \dou
\bu_B$ is not added to the forces acting on local atom B and a term
$\dou E_B^{NL} / \dou\bu_A$ is computed for atom $A$. Ghost forces don't derive
from a potential and  as a result they are not symmetrical i.~e., the ghost force on $B$ due to $A$ is not equal to that due to $A$ on $B$. Thus they
cannot be corrected in the energy directly. They are instead computed
each time the status of the representative atoms are updated and are
then held constant until next update. These dead loads $\bff^G$ can be
incorporated as corrections to the energy function: 
\bea
\Pi_h'(\bu) = \Pi_h(\bu) - \sum_\alpha \bff^G_\alpha \cdot \bu_\alpha.
\eea

The ghost forces are a function of atomic positions and thus
new ghost forces arise, once atoms relax. Their norm is linked to
the size of relaxation in transition regions. For example, in
the case of the relaxation of a twin boundary in aluminum (see
section \ref{validation}), the difference in ghost forces  before and after
relaxation is $3 \times 10^{-6}$ eV/\AA\
corresponding  to  an energy relaxation of order $10^{-7}$ eV. More
generally, we observe that the magnitude of the ghost forces is
decreased only by a factor of the order of ten when the dead load
approximation described here is applied. Nevertheless, this procedure improves the
accuracy of the solutions in transition zones.

The complete energy function including the approximate ghost force
correction is then,
\bea
\Pi_h'(\bu) = 
\sum_{e=1}^M \nu_e {\edens}(\bF_e) + 
\sum_{\beta = 1}^{\mit R_{NL}} E_\beta (\br^1_\beta,...,\br^{m_\beta}_\beta) - 
\sum_{\alpha=1}^{R} n_\alpha (\bar{\bff}_\alpha+\bff^G_\alpha)
\cdot \bu_\alpha. \label{avp_split3} 
\eea

%%%
%%%
\subsection{Energy Minimization}
\label{eneminimize}
We are interested in obtaining equilibrium configurations of the
solid. We thus invoke the principle of minimum potential energy
which states that a system will be at equilibrium when its potential
energy is minimum. The potential energy of our reduced atomic system
is given above in \eqn{avp_split3}. To minimize this energy we have
used both conjugate gradient algorithms (Papadrakakis and Ghionis 1986) 
and quasi-Newton algorithms (Denis and Schnabel 1983). The conjugate
gradient algorithm requires computation of the gradient of 
\eqn{avp_split3} with respect to the representative atom displacements 
(i.e., the system degrees of freedom). The quasi-Newton method requires
in addition also the Hessian of \eqn{avp_split3}, i.e. the second
gradient with respect to displacements. These will be evaluated in this
section for generic interatomic interactions that can be written in the form given by \eqn{EnergyDecompose}.

The gradient of the potential energy (also referred to as the 
out-of-balance force vector) is given by,
\bea
{{\partial\Pi_h'}\over{\partial\bu_\alpha}} = 
\sum_{e=1}^M \nu_e \bP(\bF_e){{\partial\bF_e}\over{\partial\bu_\alpha}} -
\sum_{\beta = 1}^{\mit R_{NL}} \left[
\sum_{j=1}^{m_\beta} \bvhi^j_\beta{{\partial\br^j_\beta}\over{\partial\bu_\alpha}}\right]
- n_\alpha (\bar{\bff}_\alpha+\bff^G_\alpha),
\label{dPu1}
\eea
where $\bP=\partial\edens/\partial\bF$ is the first Piola-Kirchhoff stress
tensor and $\bvhi^j_\beta=-\partial E_\beta/\partial\br^j_\beta$ is the
force on atom $\beta$ due to its neighbor $j$. The deformation gradient $\bF_e$
can be expressed in terms of the nodal displacements and finite element
interpolation functions as,
\bea
\bF_e=\bI+\sum_{\alpha=1}^R\bu_\alpha\nabla_0 N_\alpha(\bX_e),
\label{Fe}
\eea
where $\nabla_0=\partial/\partial\bX$ is the material deformation gradient, $\bI$ is the
identity matrix, and for the constant strain elements we use, $\bX_e$ is the
element centroid. We then have,
\bea
{{\partial\bF_e}\over{\partial\bu_\alpha}}=\nabla_0 N_\alpha(\bX_e),
\label{dFu}
\eea
and similarly by using \eqn{rjb},
\bea
{{\partial\br^j_\beta}\over{\partial\bu_\alpha}}=
\left(N_\alpha(\bX^j_\beta)-\delta_{\alpha \beta}\right)\bI.
\label{dru}
\eea
Substituting eqs.~(\ref{dFu})-(\ref{dru}) into \eqn{dPu1} and rearranging
we have,
\bea
{{\partial\Pi_h'}\over{\partial\bu_\alpha}} = 
\sum_{e=1}^M \nu_e \bP(\bF_e)\nabla_0 N_\alpha(\bX_e)-
\sum_{\beta = 1}^{\mit R_{NL}} \left[
\sum_{j=1}^{m_\beta} \bvhi^j_\beta N_\alpha(\bX^j_\beta)\right]+
\sum_{j=1}^{m_\alpha}\bvhi^j_\alpha
-n_\alpha (\bar{\bff}_\alpha + \bff^G_\alpha),
\label{dPu2}
\eea
where it is understood that the third sum is zero when atom $\alpha$ is local.
In the limit where all atoms are local the ghost force contribution drops out 
and \eqn{dPu2} reduces to,
\bea
\left.{{\partial\Pi_h'}\over{\partial\bu_\alpha}}\right|_{\mit loc}=
\sum_{e=1}^M \nu_e \bP(\bF_e)\nabla_0 N_\alpha(\bX_e)-n_\alpha \bar{\bff}_\alpha.
\label{dPu_local}
\eea
We thus have a continuum representation of the boundary value problem with
the exception that the constitutive law, $\bP(\bF)=\partial\edens/\partial\bF$,
is nonlinear and obtained from an atomistic calculation.

In the fully-refined limit where all atoms are represented and nonlocal,
the ghost forces similarly drop out, and so does the term with the
local contributions. We can
separate the first nonlocal sum into two parts,
\bea
\left.{{\partial\Pi_h'}\over{\partial\bu_\alpha}}\right|_{\mit nonloc}=
-\sum_{\stackrel{\beta=1}{\beta\neq \alpha}}^{\mit R_{NL}} \left[
\sum_{j=1}^{m_\beta} \bvhi^j_\beta N_\alpha(\bX^j_\beta)\right]
-\sum_{j=1}^{m_\alpha}\bvhi^j_\alpha N_\alpha(\bX^j_\alpha)
+\sum_{j=1}^{m_\alpha}\bvhi^j_\alpha-\bff_\alpha.
\label{dPu_nonl1}
\eea
Here the shape function acts as a Kronecker delta function, thus $N_\alpha(\bX^j_\beta)$
is equal to one when neighbor $j$ of atom $\beta$ is atom $\alpha$. We thus see that
the second sum drops out since $N_\alpha(\bX^j_\alpha)$ is always zero
(atom $\alpha$ cannot be a neighbor of itself), so \eqn{dPu_nonl1} reduces to
\bea
\left.{{\partial\Pi_h'}\over{\partial\bu_\alpha}}\right|_{\mit nonloc}=
-\sum_{\stackrel{\beta=1}{\beta\neq \alpha}}^{\mit R_{NL}} \left[
\sum_{j=1}^{m_\beta} \bvhi^j_\beta N_\alpha(\bX^j_\beta)\right]
+\sum_{j=1}^{m_\alpha}\bvhi^j_\alpha-\bff_\alpha.
\label{dPu_nonl2}
\eea
The first term is the sum of forces exerted on all other atoms by atom $\alpha$.
The second term is the sum of forces exerted on atom $\alpha$ by all of its
neighbors. The third term is the external force acting on atom $\alpha$.

For conjugate gradient minimization the expression in \eqn{dPu2} is all that
is needed. The algorithm constructs a series of conjugate search directions 
from the current gradient and previous search directions and proceeds to 
minimize along each direction using a line search routine (see Papadrakakis 
and Ghionis (1986) for more details). An alternative approach is to iteratively
solve $\partial\Pi_h'/\partial\bu_\alpha=0$ by substituting,
\bea
\left.{{\partial\Pi_h'}\over{\partial\bu_\alpha}}\right|_{I+1}\approx
\left.{{\partial\Pi_h'}\over{\partial\bu_\alpha}}\right|_{I}+
\sum_\beta \left.{ {\partial^2\Pi_h'}\over{\partial\bu_\alpha \partial\bu_\beta}}\right|_{I}
\delta\bu_\beta^{I+1}=0,
\label{taylor}
\eea
where $I$ is an iteration counter. The Hessian or second gradient expression
is the stiffness matrix $\bK_{\alpha \beta}$ and is obtained by differentiating
\eqn{dPu2}. Following a similar procedure to that used for obtaining
\eqn{dPu2} itself we have,
\bea
\bK_{\alpha \beta}=\sum_{e=1}^M\nu_e\bC(\bF_e)\nabla_0 N_\alpha(\bX_e)\nabla_0 N_\beta(\bX_e)+
\sum_{\gamma=1}^{\mit R_{NL}}\left[
\sum_{k=1}^{m_\gamma}\sum_{l=1}^{m_\gamma}
\bkap_\gamma^{kl}N_\alpha(\bX_\gamma^k)N_\beta(\bX_\gamma^l)\right] \nonumber \\
-\sum_{k=1}^{m_\alpha}\sum_{l=1}^{m_\alpha}\bkap_\alpha^{kl}N_\beta(\bX^l_\alpha)
-\sum_{k=1}^{m_\beta}\sum_{l=1}^{m_\beta}\bkap_\beta^{kl}N_\alpha(\bX^k_\beta)
+\delta_{\alpha \beta}\sum_{k=1}^{m_\alpha}\sum_{l=1}^{m_\alpha}\bkap_\alpha^{kl},
\eea
where $\bC=\partial^2\edens/\partial\bF^2$ is the Lagrangian tangent
stiffness tensor and $\bkap_\beta^{kl}=\partial^2 E_\beta/
\partial\br^k_\beta\partial\br^l_\beta$ is an atomic level stiffness.

The solution then proceeds iteratively by solving \eqn{taylor} as stated
or in reduced notation,
\bea
\sum_\beta \bK_{\alpha \beta}^I \delta\bu_\beta^{I+1}+\bF_\alpha^I=0,
\eea
where $\bF_\alpha^I=(\partial\Pi_h'/\partial\bu_\alpha)|_I$ is the out-of-balance
force vector defined above in \eqn{dPu2}. In the Newton-Raphson method
the displacements at each iteration are updated by,
\bea
\bu_\alpha^{I+1}=\bu_\alpha^I+\delta\bu_\alpha^{I+1},
\eea
while for a quasi-Newton solver a line search minimization is done along
the search directions given by $\delta\bu_\alpha^{I+1}$. 
The procedure continues until $||\bF_\alpha^I||$ is reduced
sufficiently for all $\alpha$.

%%%
\subsection{Automatic Adaption}
\label{adaption}
The realization that much of the computation in straightforward
atomistic simulation is wasted due to the sufficiency of local
continuum approximations far from defects is not new. A number of
mixed continuum and atomistic models have been proposed in recent
years to capitalize on this feature (some were referenced in the
introduction and others can be found in Tadmor 1996). The standard
approach in these models is to {\it a priori} identify the
atomistic and continuum regions and tie them together with
some appropriate boundary conditions. In addition to the disadvantage
of introducing artificial numerical interfaces into the problem, a
further drawback of these models is their inability to adapt to
changes in loading and an evolving state of deformation.  Take for example the
problem of nanoindentation. As the loading progresses
and dislocations are emitted  under the indenter,
the computational model must be able to adapt and change in accordance
with these new circumstances.  

In the current formulation we tie the need for automatic adaption to
an estimate of the error introduced by the reduction of degrees of
freedom.  It is then possible to identify regions where this error
estimator is high, and subsequently add degrees of freedom in these
regions.  The result is an automatic adaption scheme analogous to
adaptive remeshing in finite elements.

To include such an adaption procedure in 
the method, we appeal to  finite element literature, where error estimators
and automatic mesh refinement have been subjects of extensive research.
Recall that our collection of representative atoms are also nodes on a 
finite element mesh of constant strain triangles.  Thus, we use the error
estimator first introduced by Zienkiewicz and Zhu (1987) in terms of 
stresses and later modified by Belytschko and Tabbara (1993) to estimate errors
in the strain fields.  In our case, the deformation gradient, $\bF$ is
already needed for computing atomic energies, forces and stiffness, and 
therefore it is convenient to write the Zienkiewicz-Zhu error estimator
directly in terms of the deformation gradient.  Thus, we define the 
discretization error in element $e$ as 
\be
\epsilon_e=\Biggl[{{1}\over{\Omega_e}}\int_{\Omega_e}(\bar{\bF}-\bF_e)^T
(\bar{\bF}-\bF_e)d\Omega
     \Biggr]^{1/2},
\label{adapcr}
\ee
where $\bF_e$ is the finite element solution for the deformation 
gradient in element $e$, and $\bar{\bF}$ is the 
$L_2$-projection of the finite element solution for $\bF$, given by 
\be
\bar{\bF}={\bf Nf}.
\label{l2proj}
\ee
Here, ${\bf N}$ is the shape function array, and ${\bf f}$ is the
array of nodal values of the projected deformation gradient
$\bar{\bF}$.  Because the deformation gradient is constant within each
constant strain element, the nodal values ${\bf f}$ are simply
computed by averaging the deformation gradients over all of the elements
in contact with the node of interest.  The integral in equation \eqn{adapcr} can be
computed quickly and accurately using a three-point Gaussian
quadrature rule.  Elements for which the error $\epsilon_e$ is
greater than some prescribed error tolerance $\epsilon_A$ are targeted for
refinement.

Refinement then proceeds by adding three new
representative atoms at the atomic sites closest to the midsides of
the targeted elements.  
Notice that since representative atoms must
fall on actual atomic sites in the reference volume $B_0$, there is a
natural lower limit to element size.  If the nearest atomic sites to
the midsides of the elements are the atoms at the element corners, the
region is fully refined and no new representative atoms are
added. Actual examples of evolving mesh refinement is given in \fig{adex}
for the indentation problem.

In addition to mesh refinement, mesh coarsening is also an important
requirement. For example, consider the passage of a dislocation. As
the dislocation moves it leaves a trail of fully refined mesh in its
wake corresponding to previous core positions. Far behind the
dislocation the solid is undistorted and the high mesh resolution is
unnecessary and could be coarsened. To coarsen the following algorithm
is applied: (1) For each
local node/atom the elements surrounding the node and the polygon
defined by their outer sides is identified. (2) If none of these
elements satisfy the adaption criterion, remove the current local node
and create a new Delaunay triangulation of the outer polygon. (3) If
none of the new elements satisfy the adaption criterion then the local
node and all the old elements connected to it are deleted and the new
elements are accepted. Essentially, the idea is to examine the
necessity of each node. To prevent excessive coarsening of the
mesh far from defects, the nodes corresponding to the initial mesh are
usually protected from deletion.

%%%
%%%
\subsection{Putting it all together}
\begin{figure}\wheretoput
\centerline{\epsfysize=3.5truein \epsfbox{\figdir{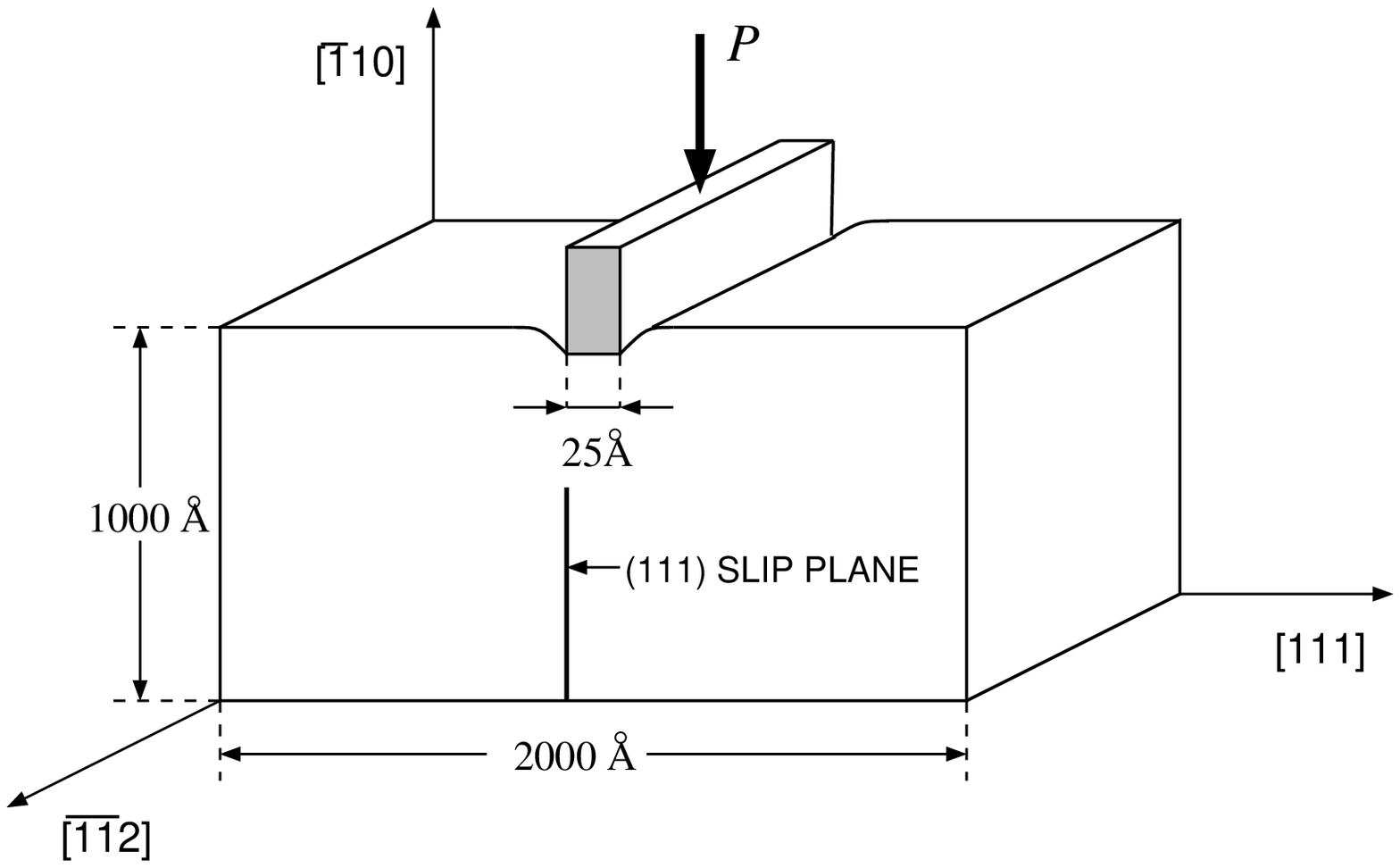}}}
\figcaption{Nanoindentation in an aluminum single crystal.}
\label{nanoprob}
\end{figure}

To demonstrate the steps involved in an adaptive quasicontinuum analysis consider
the problem of nanoindentation depicted in \fig{nanoprob}. Here a 
 rigid rectangular indenter, infinite in the out-of-plane direction, is 
 pressed into the free surface of a thin film aluminum single crystal.
The dimensions and crystallographic orientation are given in the figure. We
are interested in applying a quasicontinuum analysis to this problem (which
is too large to be comfortably tackled by direct atomistics) to obtain such
data as load versus indentation curves, the criterion for dislocation nucleation under the indenter and
stress distributions.

Different boundary conditions are possible to characterize this problem. It is
possible to model the indenter as well as the film and consider the interactions
between indenter atoms and film atoms. However in the interest of simplicity
we choose to neglect these effects and model the indenter as a rigid displacement
boundary condition, i.e. all atoms on the surface under the indenter are
forced to move down with it. In addition the displacements parallel to the
indenter face and in the out-of-plane direction can be constrained to mimic
perfect stick conditions or released for a friction free indenter. The rest
of the surface is left free and unconstrained. Far from the indenter, symmetry
boundary conditions are applied to the model right and left edges and the 
substrate is taken to be rigid with zero displacements at the interface.

We need to select an initial set of representative atoms. 
A fully-refined mesh in the vicinity of the indenter is desired from the start in order to
capture surface effects there and have sufficient resolution to accommodate the
indenter geometry. The details of the initial mesh generation may be
found in Tadmor (1996). 
\medskip
\noindent The simulation proceeds as follows:
\begin{itemize}
\item [{[1]}] Next load step -- the indenter is driven another step into the 
crystal (0.2\AA\ in these simulations) by rigidly displacing the atoms under 
the indenter downward by the appropriate amount.

\item [{[2]}] Local/nonlocal status computation -- the locality criterion 
defined in \eqn{lcrit} is evaluated for each representative atom and its
status is determined. Significant preprocessing is done at this stage (such
as the storage of atom lists and computation of shape functions) 
to speed up the computations (see Tadmor (1996) for details).

\item [{[3]}] Ghost force evaluation -- ghost forces are computed and applied 
as dead loads.

\item [{[4]}] Energy minimization -- a quasi-Newton solver is used to iteratively 
minimize the total potential energy and to identify the equilibrium configuration 
of the system subject to the new load step boundary conditions.

\item [{[5]}] Automatic adaption -- all elements in the mesh satisfying the adaption
criterion  are adapted, i.e., divided into smaller 
elements. Since all nodes must occupy atomic sites a natural cutoff commensurate
with the lattice spacing prevents indefinite adaption. At the same time that
elements are being checked for adaption they can also be checked for coarsening
and removed if they are not necessary.

\item [{[6]}] If elements have been added (or removed) in the adaption phase the system 
will no longer be in equilibrium (since the system has changed). In this case
return to [2] to obtain the new relaxed configuration.

\item [{[7]}] Output -- load/displacement data, displacement contours, stress and strain
contours, energy, atomic structure, etc.

\item [{[8]}] Proceed to [1] for next load step.
\end{itemize}

\begin{figure}\wheretoput
\centerline{\epsfysize=3.0truein \epsfbox{\figdir{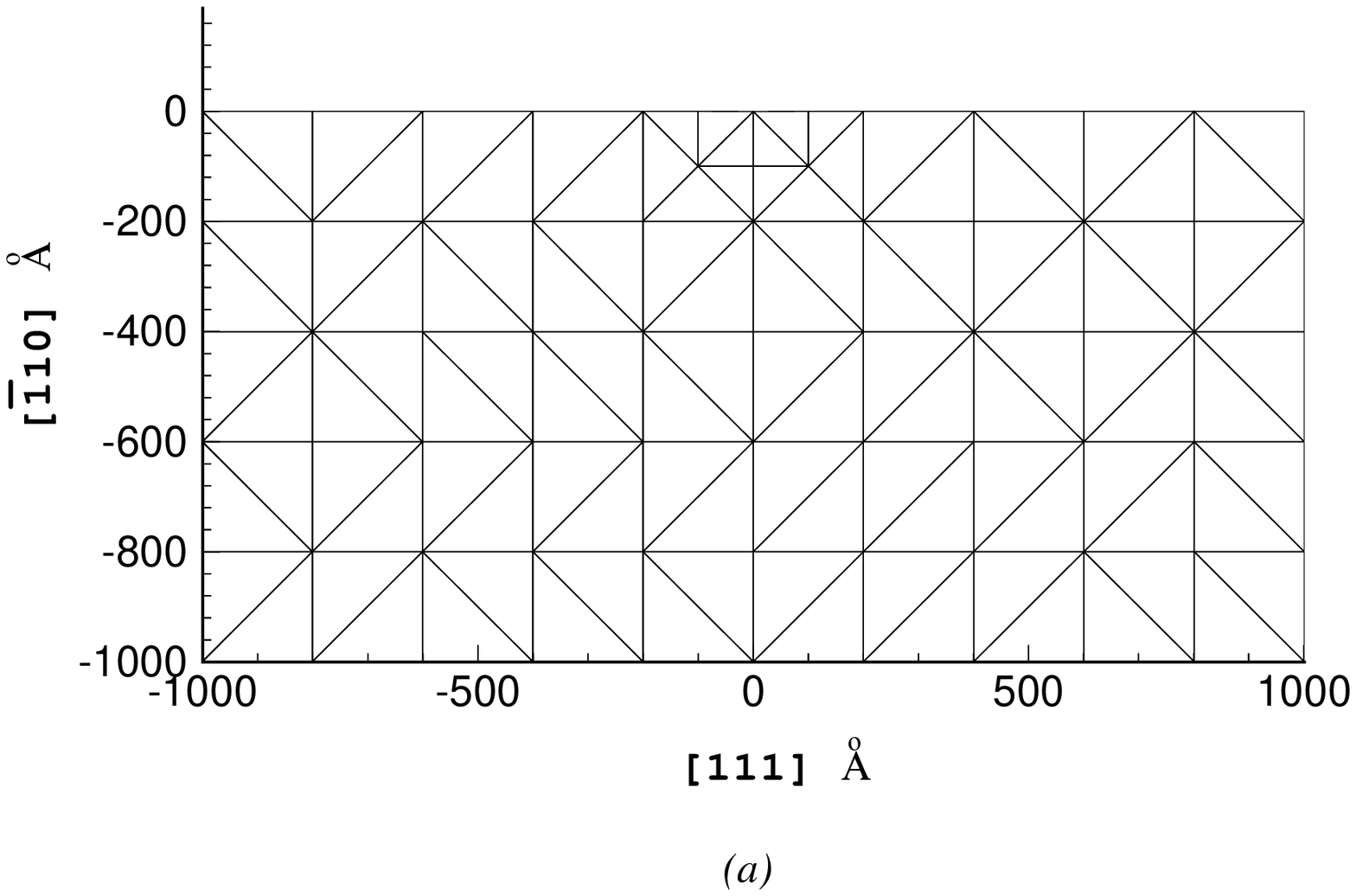}}}
\centerline{\epsfysize=3.0truein \epsfbox{\figdir{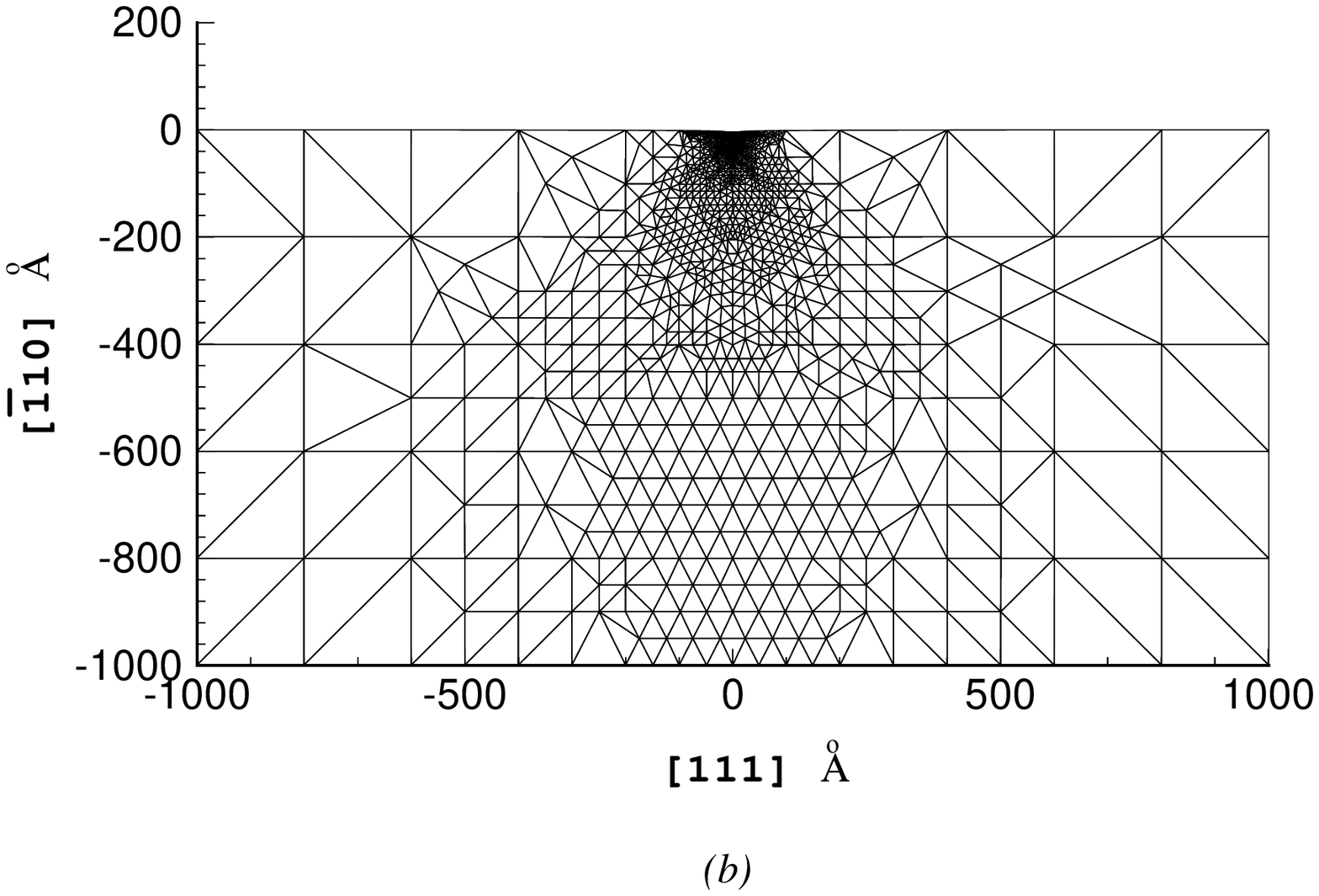}}}
\centerline{\epsfysize=3.0truein \epsfbox{\figdir{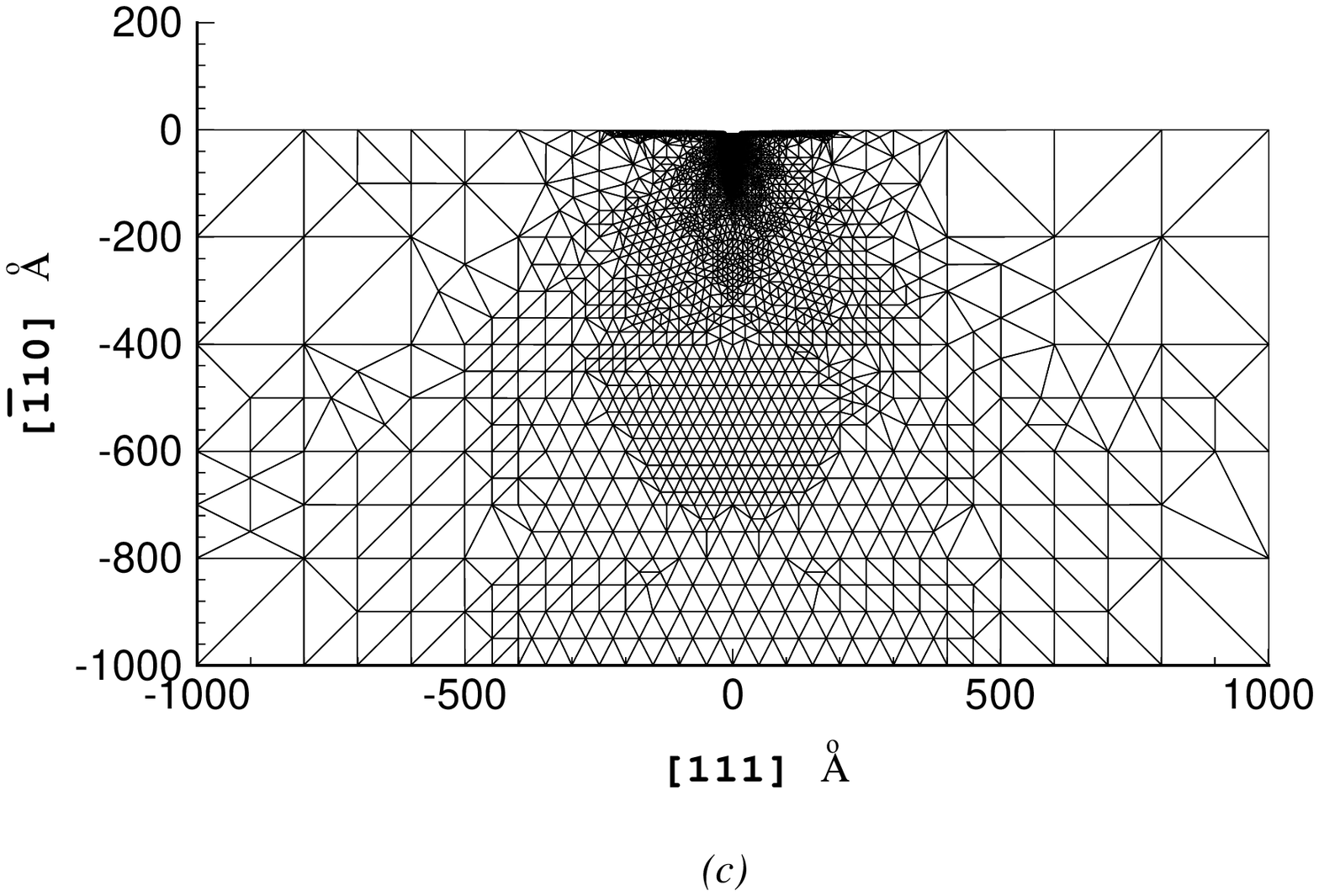}}}
\figcaption{Automatic adaption process in action for the problem of nanoindentation.}
\label{adex}
\end{figure}

\begin{figure}\wheretoput
\centerline{\epsfysize=2.5truein \epsfbox{\figdir{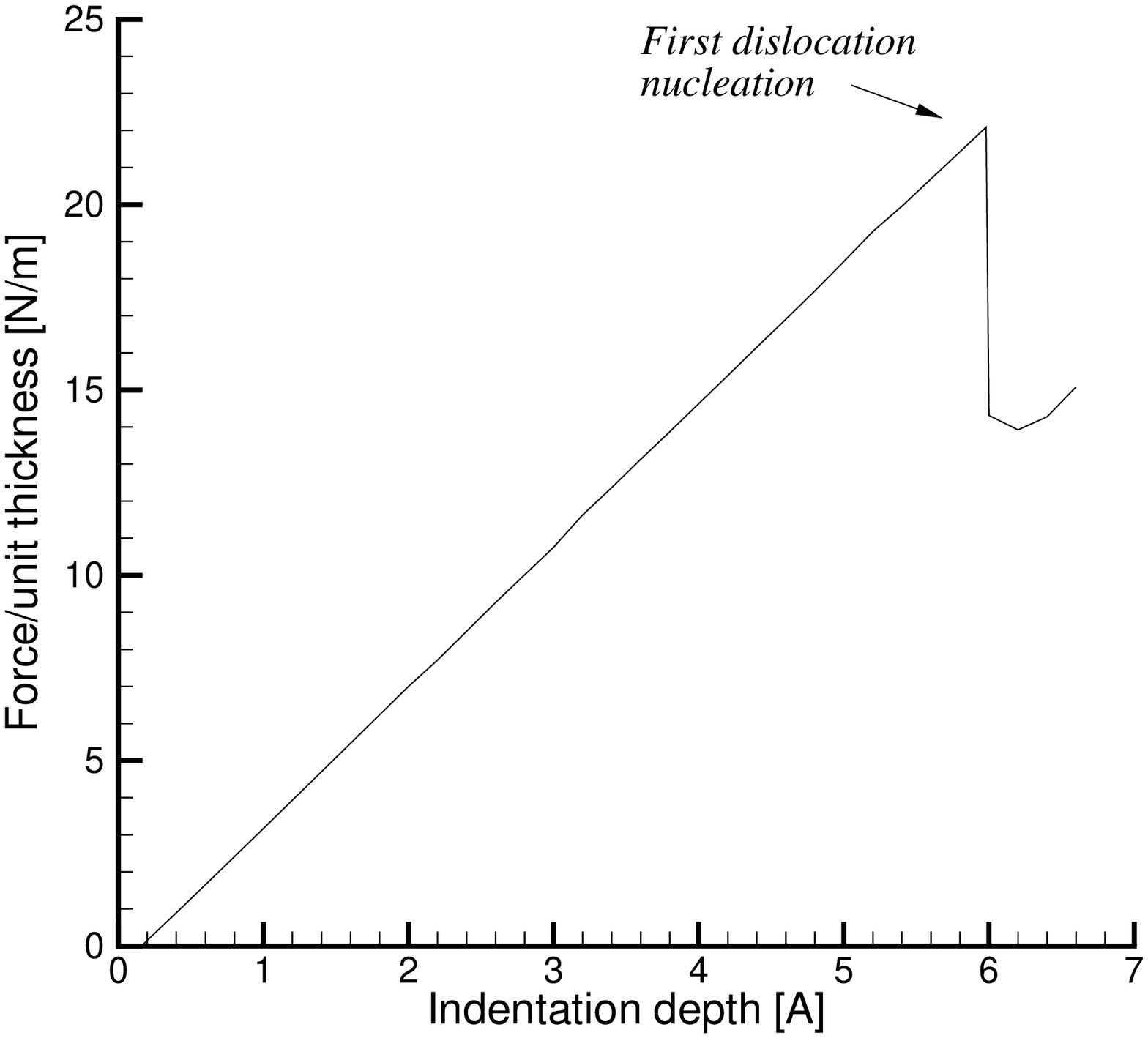}}\epsfysize=2.5truein \epsfbox{\figdir{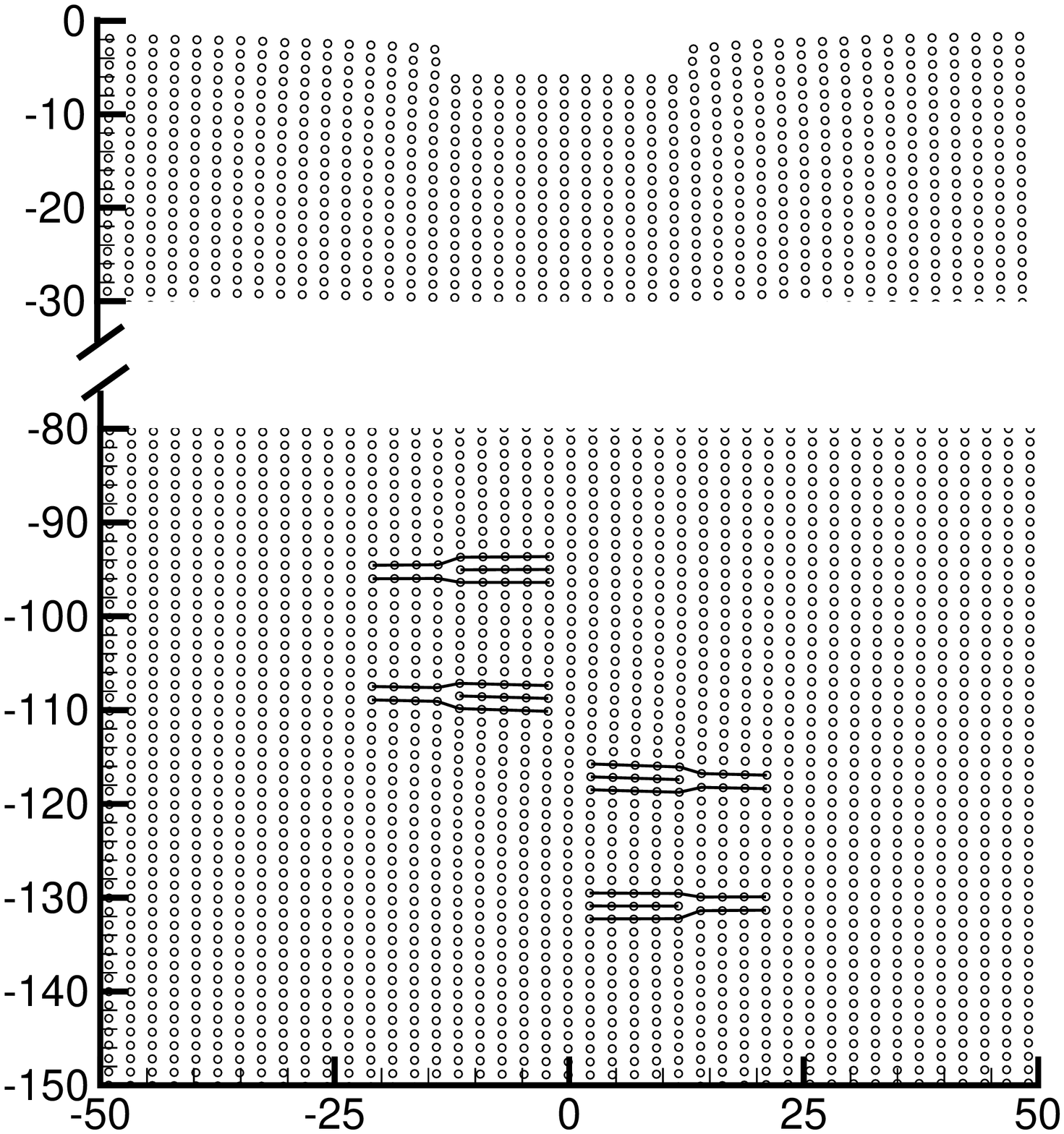}}}
\centerline{ \hspace{0.55truein} (a) \hspace{2.2truein} (b)}
\figcaption{(a) Load vs.~indenter displacement. Result obtained from QC
simulation of nanoindentation on a single crystal Al. (b) Atomic structure under indenter after nucleation  of a dislocation. }
\label{nanoresult}
\end{figure}

A series of snapshots of the progressing adaption for this problem were given
in \fig{adex}. The load-displacement curve computed for an
embedded atom model of aluminum due to Ercolessi and Adams (1992) is given
in \fig{nanoresult}a. The response is initially linear as predicted
by elasticity theory, until dislocations are nucleated at a critical load.
The atomic structure under the indenter after dislocation nucleation is 
presented in \fig{nanoresult}b. A far more detailed discussion of this
simulation and others for different orientations and indenter geometries
are presented in Tadmor et al.~(1997).

In section \ref{nanogb} nanoindentation in an aluminum bicrystal is discussed
in more detail. There the nanoindentation is used as a means for generating
dislocations, as a ``dislocation gun'', in order to probe the interaction of 
dislocations with grain boundaries.

%%%
%%%
%%%
\section{Validation} 
\label{validation}

As was noted in earlier work, our aim with the quasicontinuum method
is to properly recover two limiting cases.  On the one hand, one aims
to restore conventional atomistic simulation in the limit that full
atomic resolution is adopted everywhere.  The benchmark of such a
success is whether or not the quasicontinuum results for defect cores
are in accord with those obtained by conventional atomistic simulation.
On the other hand, restoration of the continuum limit in the event of
only long wavelength deformations is revealed in features such as the
appropriate dispersion relation for long wavelength elastic waves.

This section contains the validation of the formulation presented in
the previous section. We compare the results obtained using QC with
those obtained from lattice statics (LS) in the cases of a (111) free
surface in aluminum, a twin boundary in aluminum, a $\Sigma 5$
boundary in gold, and a $\Sigma 99$ boundary in aluminum.  Aluminum
was modeled using the embedded-atom potentials developed by Ercolessi
and Adams (1993) and gold with the Finnis-Sinclair potentials of
Ackland, Tichy, Vitek, Finnis (1987). In all the cases, representative
atoms within a distance of $2 \rcut$ from the interface are treated using
the nonlocal rule.

The (111) free surface in aluminum was modeled using a block of atoms with dimension 
$114 \mbox{\AA} \times 65 \mbox{\AA}$. 
\Tab{SurfTable} shows a comparison of the
relaxed energies of the atoms in different layers with
those obtained from direct atomistics.  The energies of the first three
layers are in good agreement with those computed via
LS.  The relaxation process brings about a separation of $0.021 \mbox{\AA}$
between layers $A$ and $B$ and the layers $B$ and $C$ are closer by
$0.003 \mbox{\AA}$ which are equal to those obtained with LS.

\begin{figure}\wheretoput
\centerline{\epsfysize=3.5truein \epsfbox{\figdir{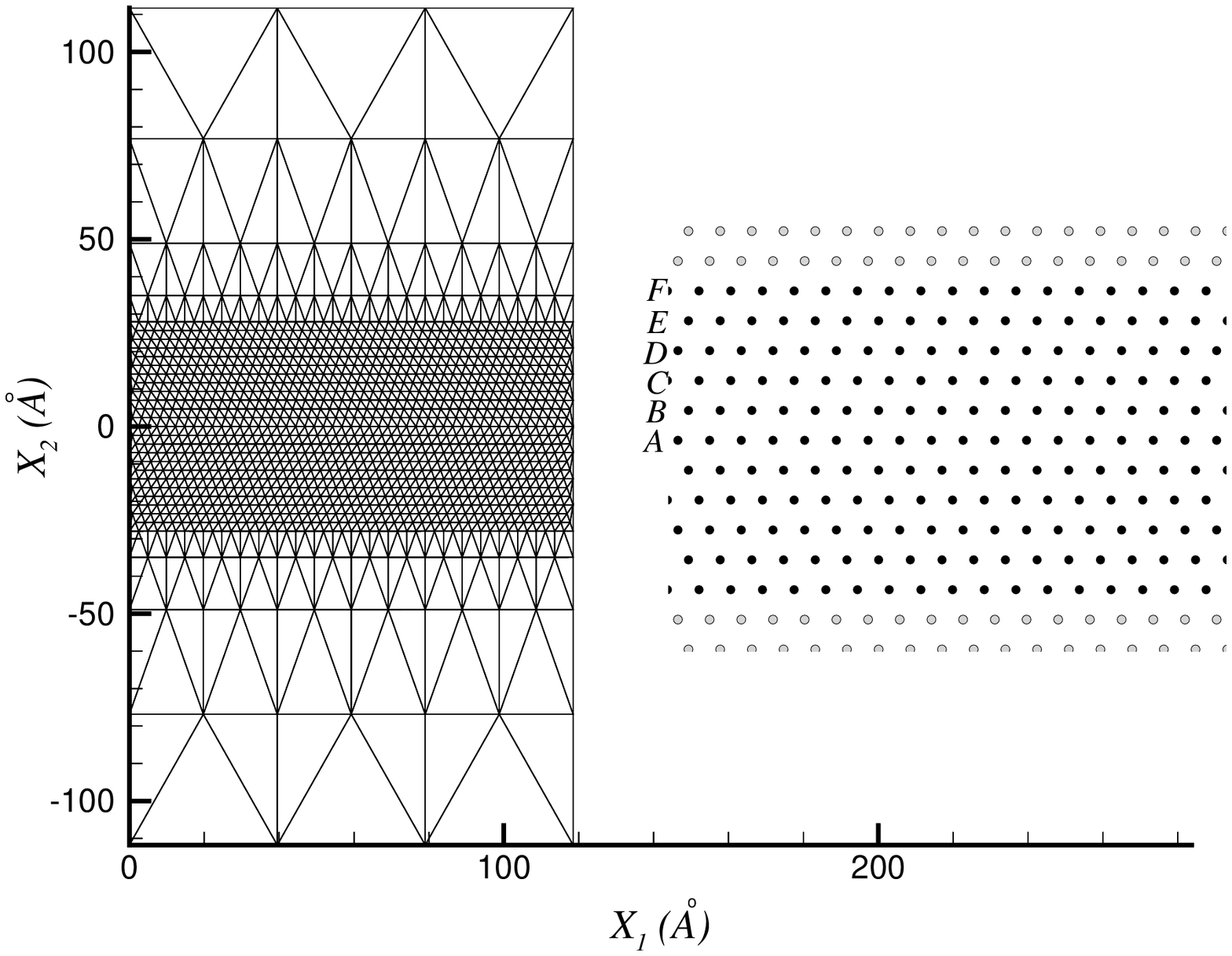}}}
\figcaption{Finite element mesh used in the simulation of a twin boundary
in Al. Inset shows structure of the boundary. $\bullet$--nonlocal representative atoms, $\circ$--local representative atoms}
\label{twin} 
\end{figure}

\begin{figure}\wheretoput
\centerline{\epsfysize=4.0truein \epsfbox{\figdir{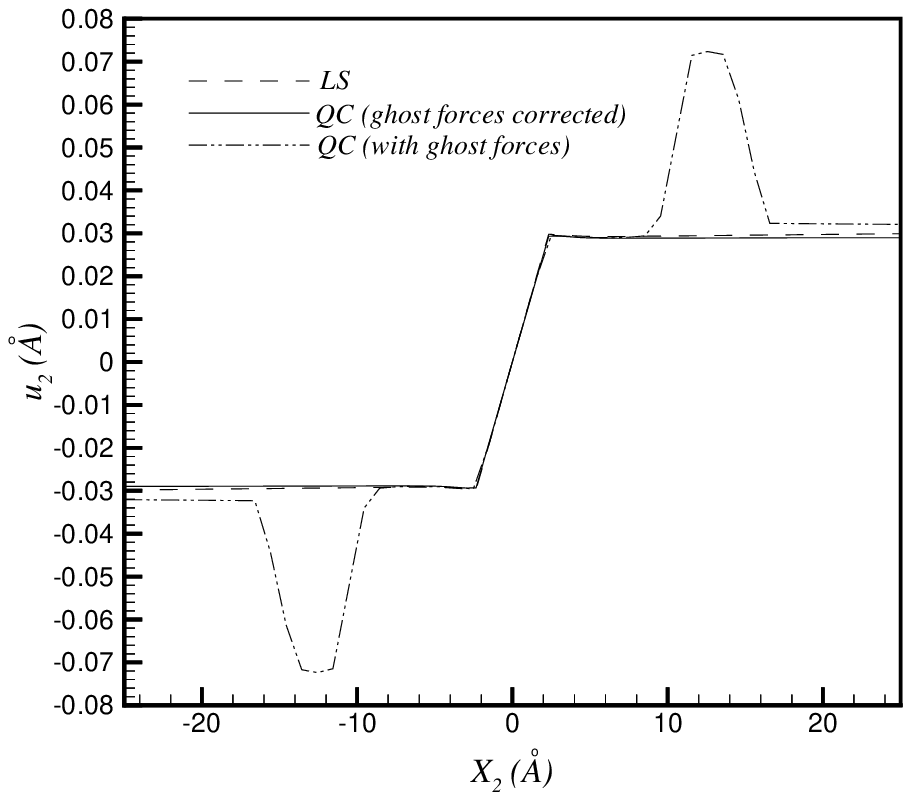}}}
\figcaption{Comparison of the QC solution (with and without the ghost
force correction algorithm) with LS for the atomic displacements in the
vicinity of a twin boundary in Al.}
\label{twcomp}
\end{figure}

A block of size 118\AA $\times$ 222\AA~was used to simulate a twin
boundary ($\Sigma = 3 (11\bar{2})$) in aluminum. The twin plane was chosen to 
be the $y=0$ plane
as shown in \fig{twin}. The comparison of relaxed energies  obtained
from LS and QC are shown in \tab{TwinTable}. The $y$-displacement ($u_2$) obtained from QC (with  and without the ghost force
removal algorithm)  is compared with that from LS in
\fig{twcomp}, where the agreement is seen to be excellent when the
ghost forces are corrected. The errors
caused by the ghost forces are also seen in this figure.

\begin{figure}\wheretoput
\centerline{\epsfysize=3.0truein \epsfbox{\figdir{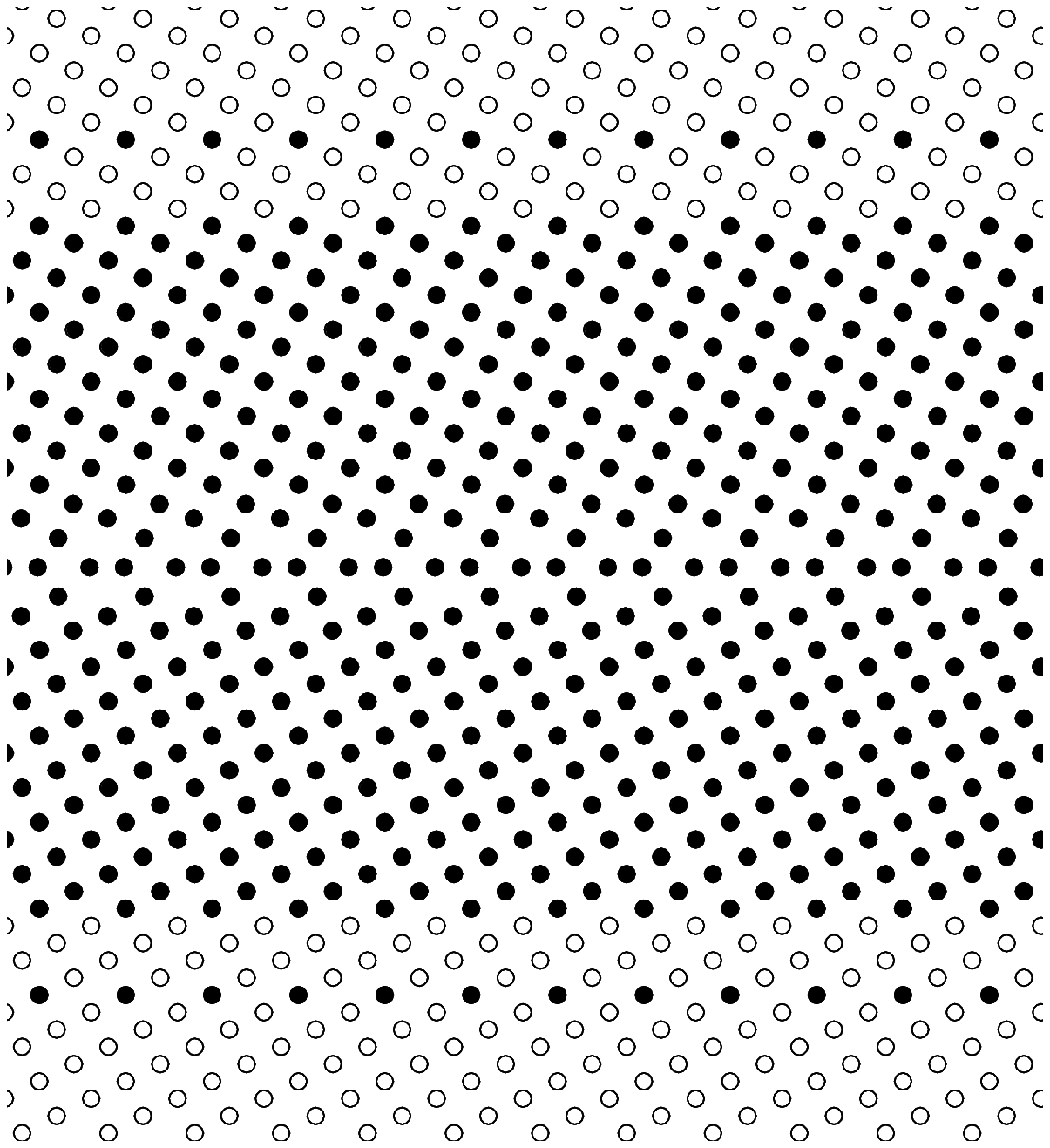}} \epsfysize=3.0truein \epsfbox{\figdir{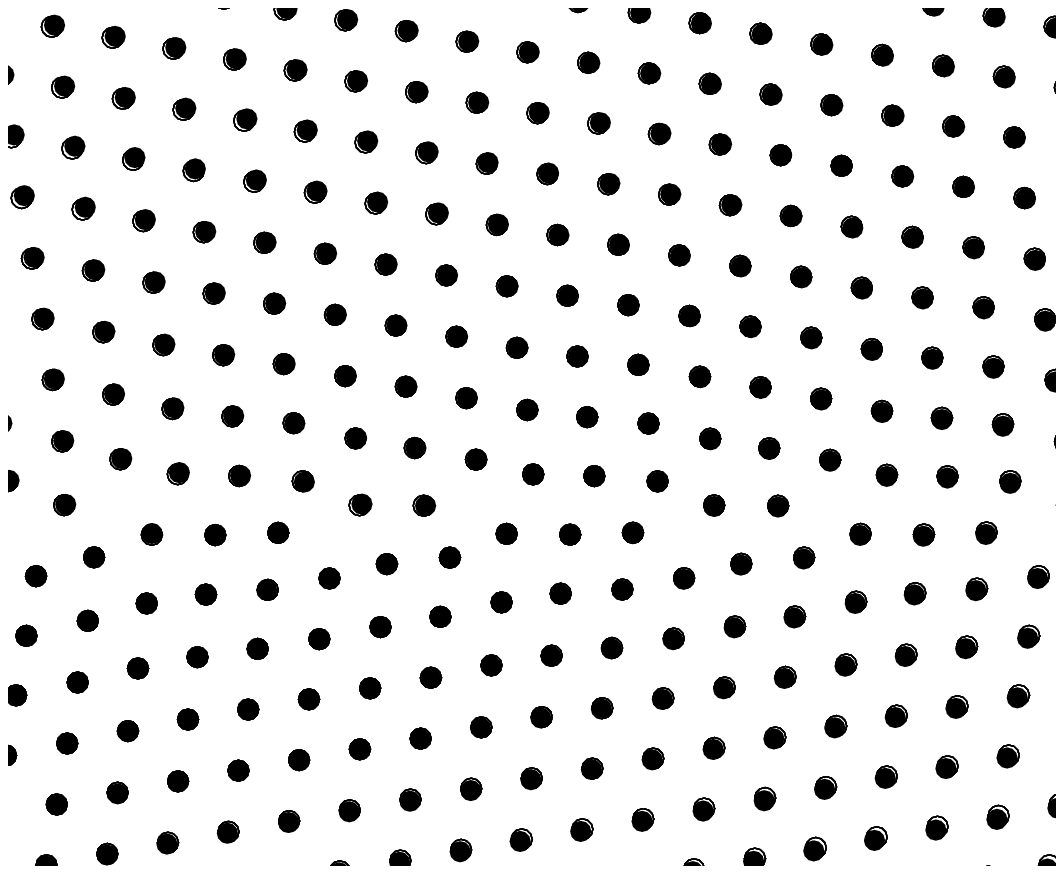}}}
\centerline{ (a)  \hspace{3.truein} (b)}
\figcaption{(a)Comparison of atomic structure  of  a $\Sigma 5 (210)$ boundary in
Au computed using the QC method with that obtained from
LS. (b) Comparison of atomic structure of a $\Sigma 99 (557)$  boundary in
Al computed using the QC method with that obtained from
LS.  $\circ$--LS, $\bullet$--representative atoms in QC}
\label{sigbds}
\end{figure}

Ackland, Tichy, Vitek and Finnis (1987) have carried out lattice
statics simulations of the $\Sigma 5 (210)$ boundary in Au.
\Fig{sigbds}a shows a comparison of the structure obtained using QC
with that obtained using LS where the agreement is seen to be
excellent. The value of the grain boundary energy computed using QC is
670$mJ/m^2$ which is consonant with the 676$mJ/m^2$
obtained by Ackland, Tichy, Vitek and Finnis (1987). 

A more complex $\Sigma 99 (557)$ boundary was also simulated using QC;
the comparison with LS solution is shown in \fig{sigbds}b. The
structure obtained here also agrees well with that of Dahmen, Hetherington,
Okeefe, Westmacott, Mills, Daw and Vitek (1990) who performed a
combined theoretical/experimental study of this boundary (by visual
inspection, a quantitative comparison was not made).

We conclude this section by noting that in all the cases presented
above the QC solution of the interfacial structure agrees well with
those obtained from lattice statics. The implication of this success is that the method has been shown to be a viable alternative to lattice statics when simulating grain boundaries and thus may be used for 
simulations involving interfacial deformation.

\section{Interfacial Motion}

The macroscopic plastic behavior of a solid is a cumulative result of the
motion of dislocations.  In addition,  grain boundaries also accommodate plastic
deformation by processes such as sliding.  In such cases it is of interest
to study inhomogeneities on the grain boundary, i.e., grain boundary
defects and their interaction with an applied stress. The first
example described in this section attempts to investigate  the effect of stress
 on a twin boundary with a step. As a second example, we describe simulations
where we study dislocations interacting with grain boundaries.

\begin{figure}\wheretoput
\centerline{\epsfysize=4.0truein \epsfbox{\figdir{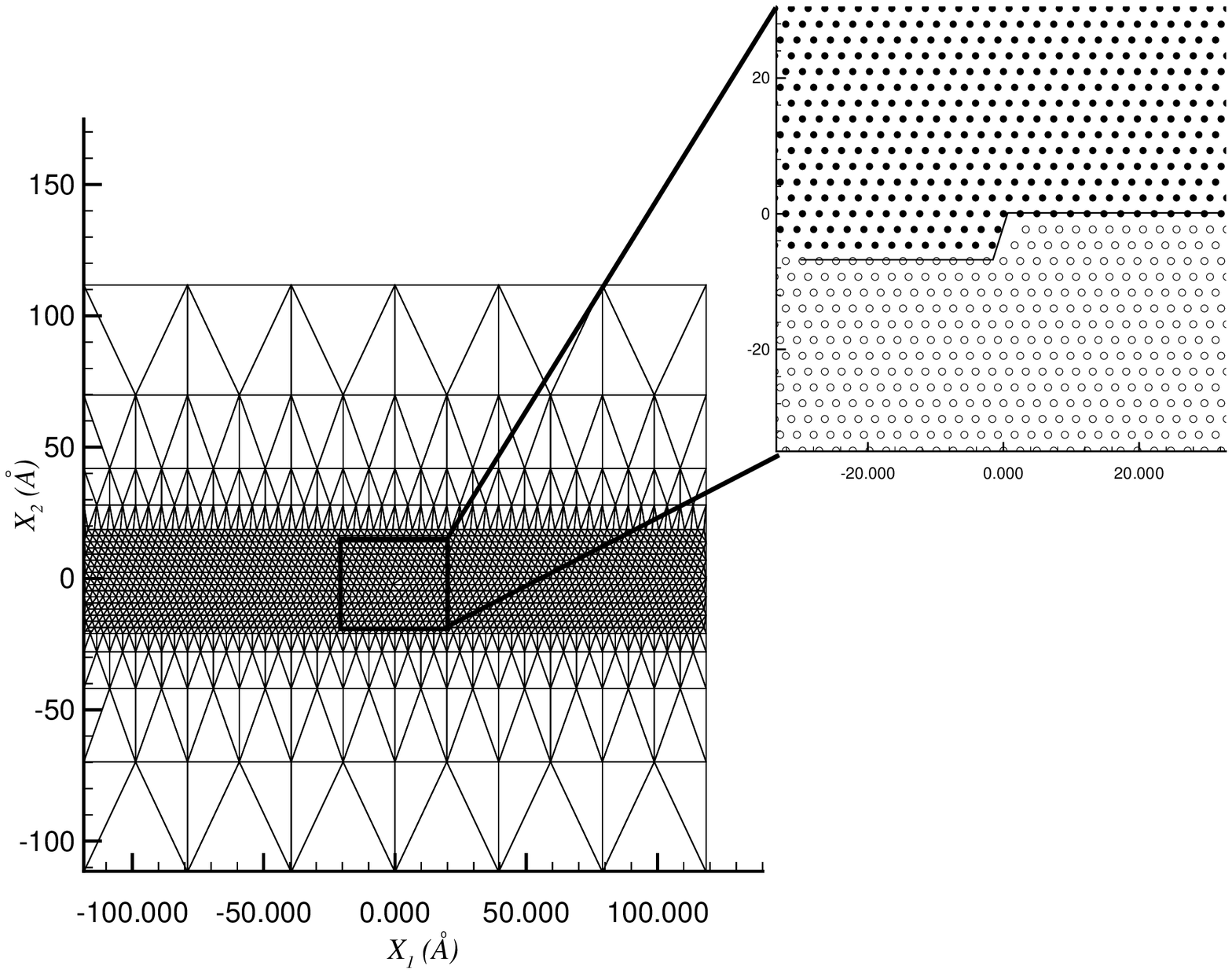}}}      
\figcaption{Finite element mesh used in the simulation of the interaction
of an applied stress with a step on a twin boundary.} \label{stepmesh}
\end{figure}

\begin{figure}\wheretoput
\centerline{\epsfysize=3.0truein \epsfbox{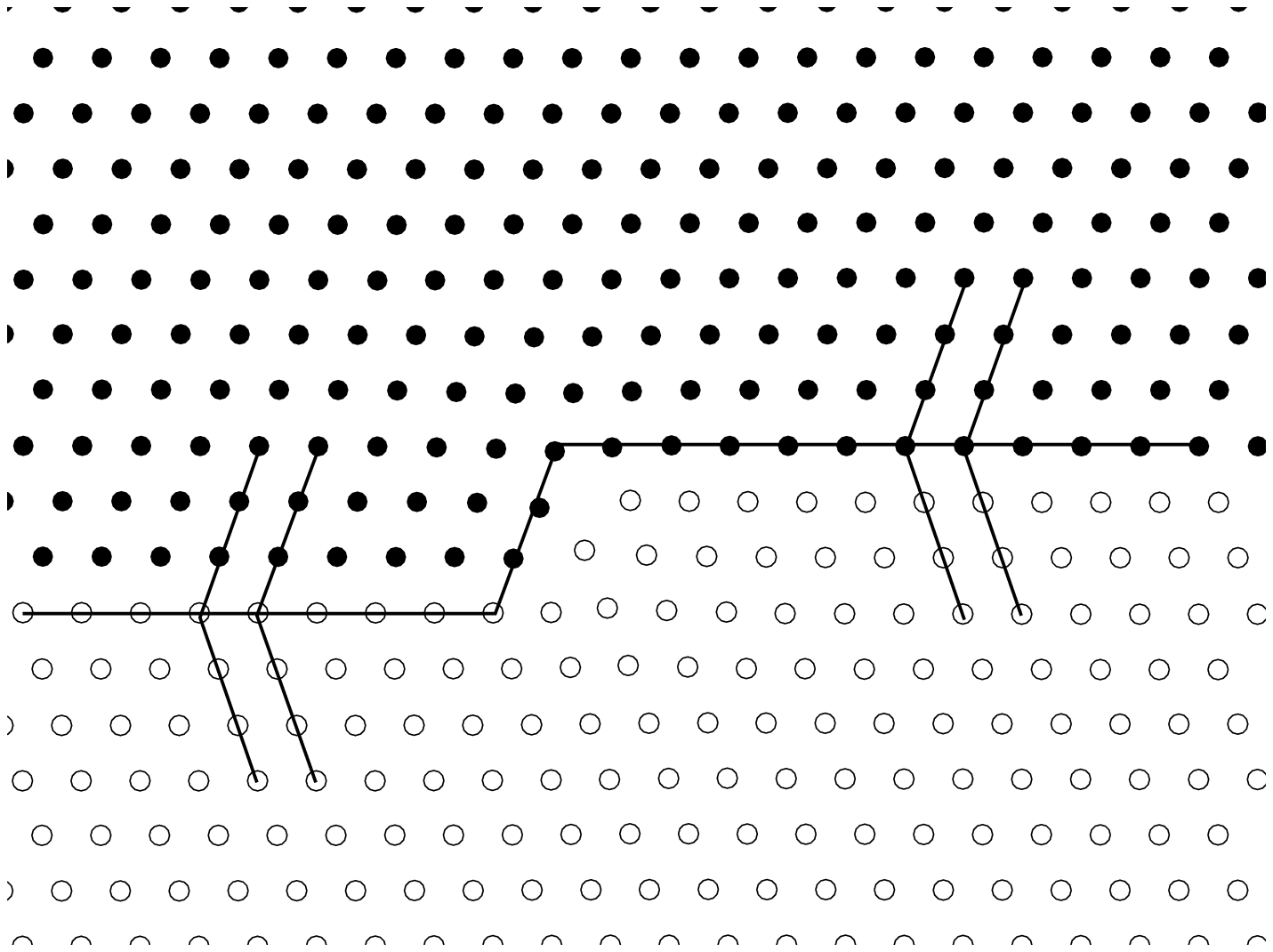} \epsfysize=3.0truein \epsfbox{\figdir{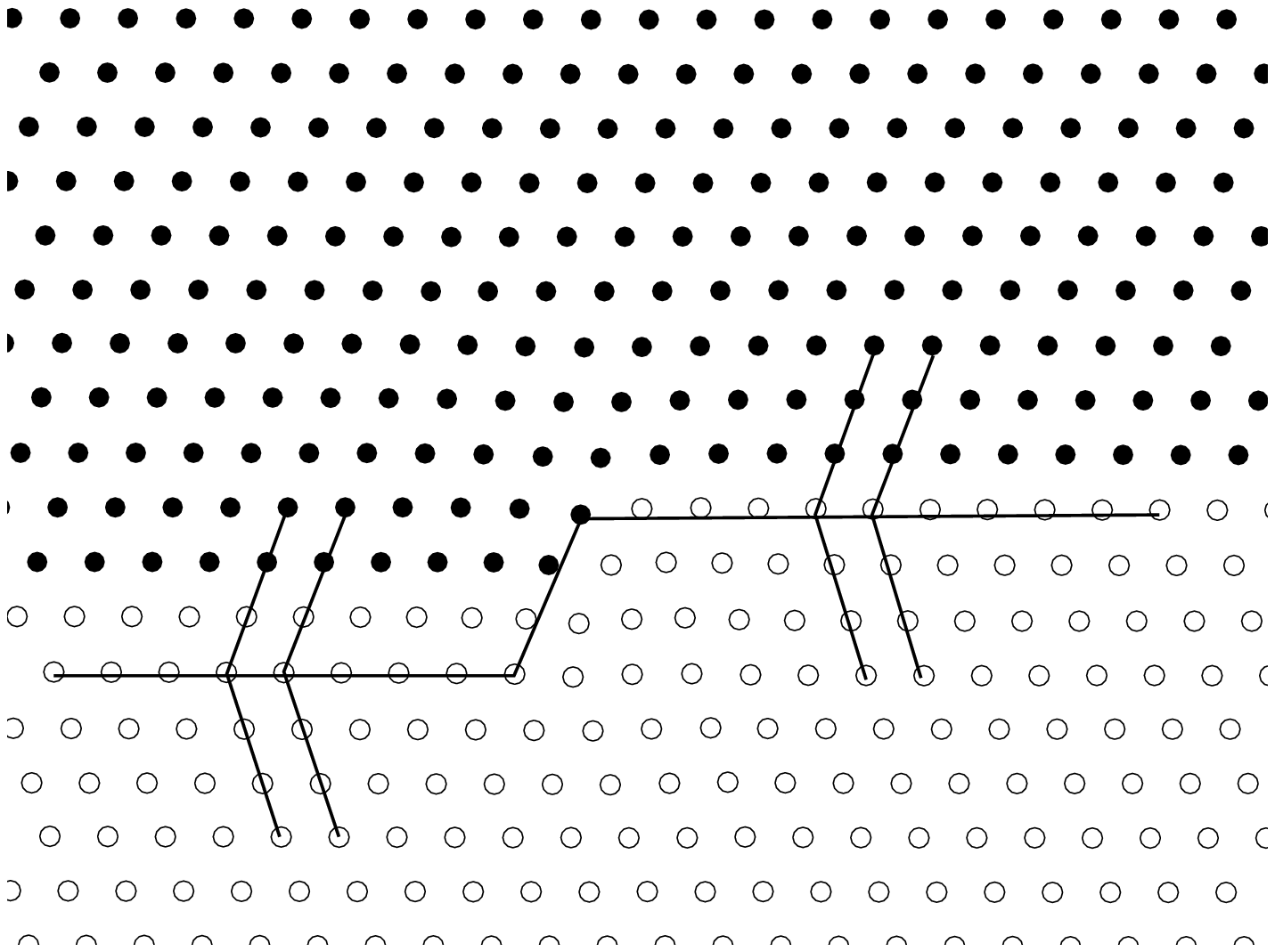}}}
\centerline{(a) \hspace{4.0truein} (b)}
\figcaption{ (a) Initial atomic configuration near the step on a twin boundary. (b) Final atomic configuration near the step; the new position of
the boundary is schematically indicated. } 
\label{stepstruct}
\end{figure}

\begin{figure}\wheretoput
\centerline{\epsfysize=4.0truein \epsfbox{\figdir{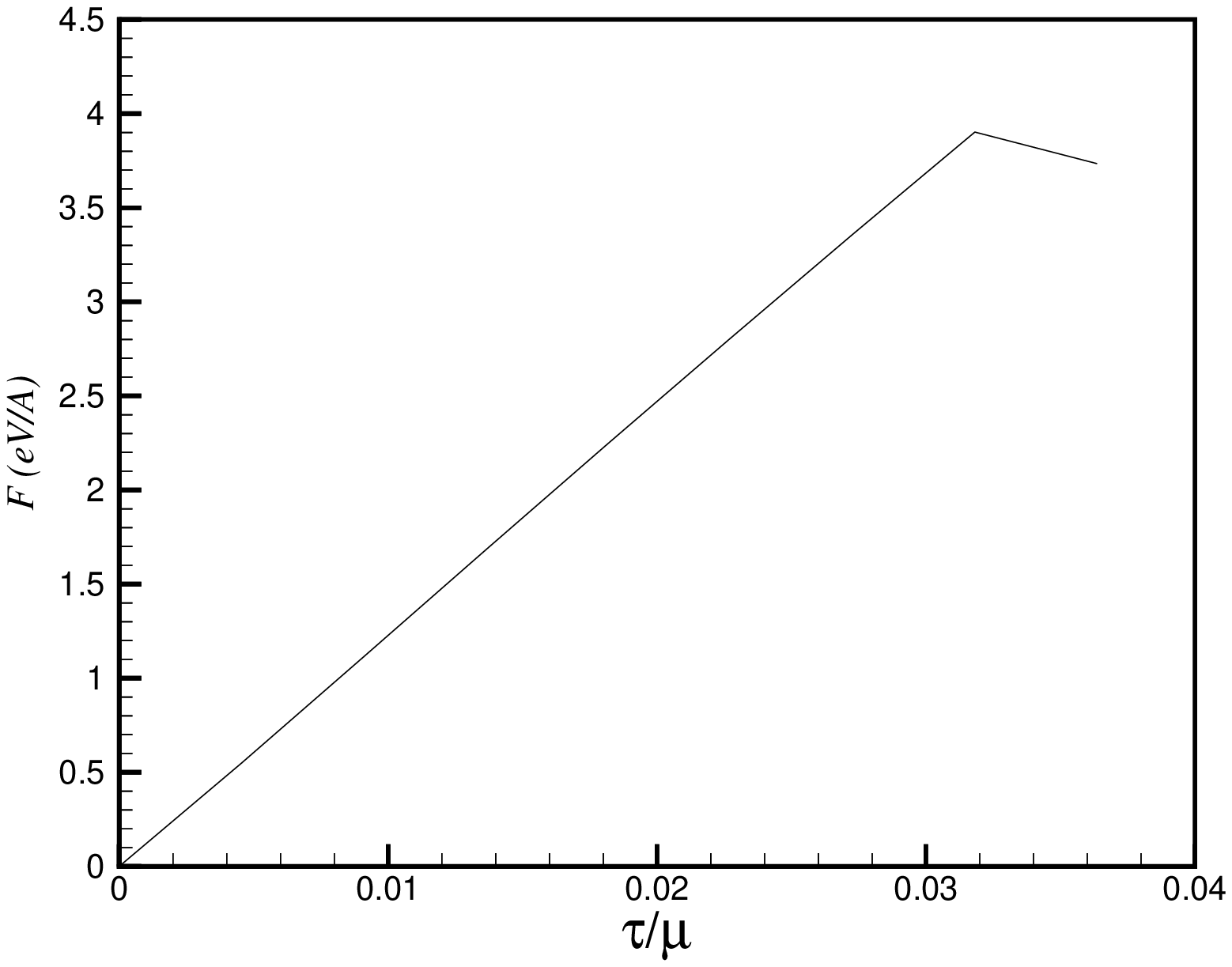}}}   
\figcaption{Load vs. applied strain response of the stepped twin
boundary. After a critical strain is reached, the step yields.} \label{loaddisp}
\end{figure}

The QC method may be used to study the interaction of interfacial
inhomogeneities with an external stress. The goals of such simulations are
twofold: (a) the determination of the critical load required to induce
plastic deformation and (b) the elucidation of the mechanism of such
deformation. The
present example is that of a step on a twin boundary ($\Sigma=3 (111)$) in
aluminum and its interaction with an applied shear stress.  The finite
element mesh and the associated step geometry are shown in  \fig{stepmesh}.
The step is subjected to a far-field homogeneous shear deformation which
is effected by the application of kinematic boundary conditions,
equivalent to the shear stress, on the boundary of the model. 
\Fig{stepstruct}a shows the relaxed configuration of the step in the
absence of applied loads. On application of the load, the solid
undergoes a near homogeneous deformation, and on attainment of a
critical stress $\tau^*$, the stepped boundary undergoes an
inhomogeneous deformation; the configuration after this event is shown
in \fig{stepstruct}b.  An examination of \fig{stepstruct} reveals the
migration of the twin boundary by the nucleation of two
$\frac{a_o}{6}[\bar{1} \bar{1} 2]$ edge dislocations from the corners
of the step.  The value of the displacement jump computed from the
simulation corresponds to the Burgers vector of a
$\frac{a_o}{6}[\bar{1} \bar{1} 2]$ dislocation which is equal to 1.646
\AA~in the case of aluminum. On nucleation, the dislocations move
towards the boundary of the simulation cell and are eventually stopped
by it.  The critical value of nondimensional stress $\tau^*/\mu$ is found to lie
between 0.031 and 0.036 and is found to be
insensitive to the cell size chosen for the simulation.  \Fig{loaddisp}
shows a plot of effective load (norm of the reaction vector) versus
the net global shear strain, where a drop in the effective load is
seen at the critical strain. This drop in the load can be estimated
using a simple linear elastic theory. It is seen that the initial part
of this curve is a linear function of the global strain, and at these
strain levels all the strain is elastic. On the nucleation of the
dislocations (at a global strain level of 0.036), the net {\em
elastic} strain falls to 0.028. The load corresponding to this strain
obtained for the linear part of the curve is 3.52 eV/\AA~while the
value of the effective load obtained from the simulation at a global
strain level of 0.036 is 3.72 eV/\AA. The value obtained from the
simulation is expected to be higher than that predicted due to the
fact that the dislocations are trapped near the boundary of the
simulation cell, and thus the elastic strain is not reduced to the
value obtained from the simple analysis.

It is interesting to contrast the critical stress $\tau^*$ with a
typical Peierls stress for a straight dislocation.  For example, in
the case of a screw dislocation in this metal the Peierls stress is
0.00068$\mu$ (Shenoy and Phillips 1997), nearly fifty times smaller
than the critical stress for advancing the twin boundary.  As another
comparison to set the scale of the stresses determined here, the
stress to induce motion of the twin boundary can be compared with that
to operate a Frank-Read source which is $\sigma \approx \mu b / L$,
where $L$ is the width of the source (Hull and Bacon 1992).  In light of
this estimate, the stress to induce motion of the twin boundary is of
the same order as that to operate a Frank-Read source of width
$\approx 35b$ (where $b$ is a typical Burgers vector). Although
typical Frank-Read sources are larger than $35b$ and hence operate at
even lower stresses, the stress found to stimulate motion of the twin
boundary is still significantly smaller than the ideal shear strength,
and is an example of the ``lubricating'' effect of
heterogeneities in the motion of extended defects.

\section{Interaction of lattice dislocations with a grain boundary}
\label{nanogb}

The interaction of dislocations with grain boundaries has been
identified as an important factor governing the yield and hardening
behavior of solids. For example, the dependence of the yield stress
on the grain size given by the celebrated Hall-Petch relationship (see, for example, Hirth and Lothe (1968)), is explained using a pile-up model which
assumes that dislocations are stopped by the grain boundary. In this
section we illustrate how the QC method can be used to build realistic
models that  address the issue of the interaction of lattice
dislocations with grain boundaries.  For the specific GB we consider,
we confirm the hypothesis that a pile-up will indeed occur, and that
no-slip transmission takes place across the boundary.

\begin{figure}\wheretoput
\centerline{\epsfysize=4.0truein \epsfbox{\figdir{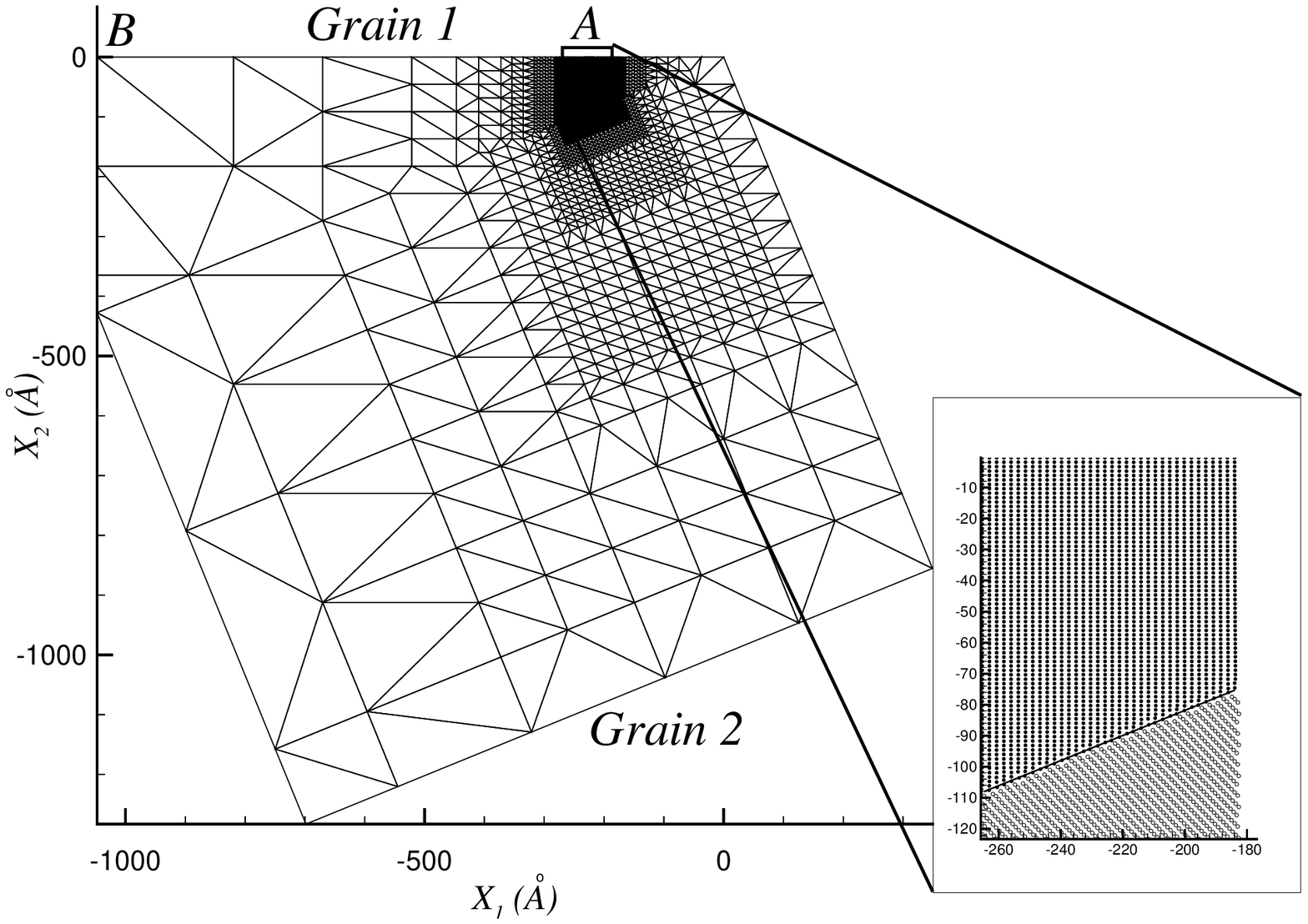}}}  
\figcaption{Mesh designed to model the interaction of dislocations and
a grain boundary. Dislocations are generated at the point $A$ by
rigidly indenting on the face $AB$ of the crystal.} \label{gindmodel}
\end{figure}

We study the interaction of $\frac{a_o}{2}[\bar{1}10]$ 
dislocations with a $\Sigma =7 (2\bar{4}\bar{1})$ symmetric tilt boundary
in aluminum. \Fig{gindmodel} shows a bicrystal, the top
face (between A and B in \fig{gindmodel}) of which is subject to a
kinematic boundary condition that mimics the effects of a rigid indenter.
On attainment of a critical load, dislocations are nucleated at the point
A, and they move towards the grain boundary.  We investigate the nature of
the interaction of these dislocations and the grain boundary through the consideration of the following questions: will the dislocation be
absorbed by the boundary, and if so what is the result of this process?
Will the dislocation cause a sufficient stress concentration at the boundary
so as to result in the nucleation of a dislocation in the neighboring
grain? 

\begin{figure}\wheretoput
\centerline{\epsfysize=3.0truein \epsfbox{\figdir{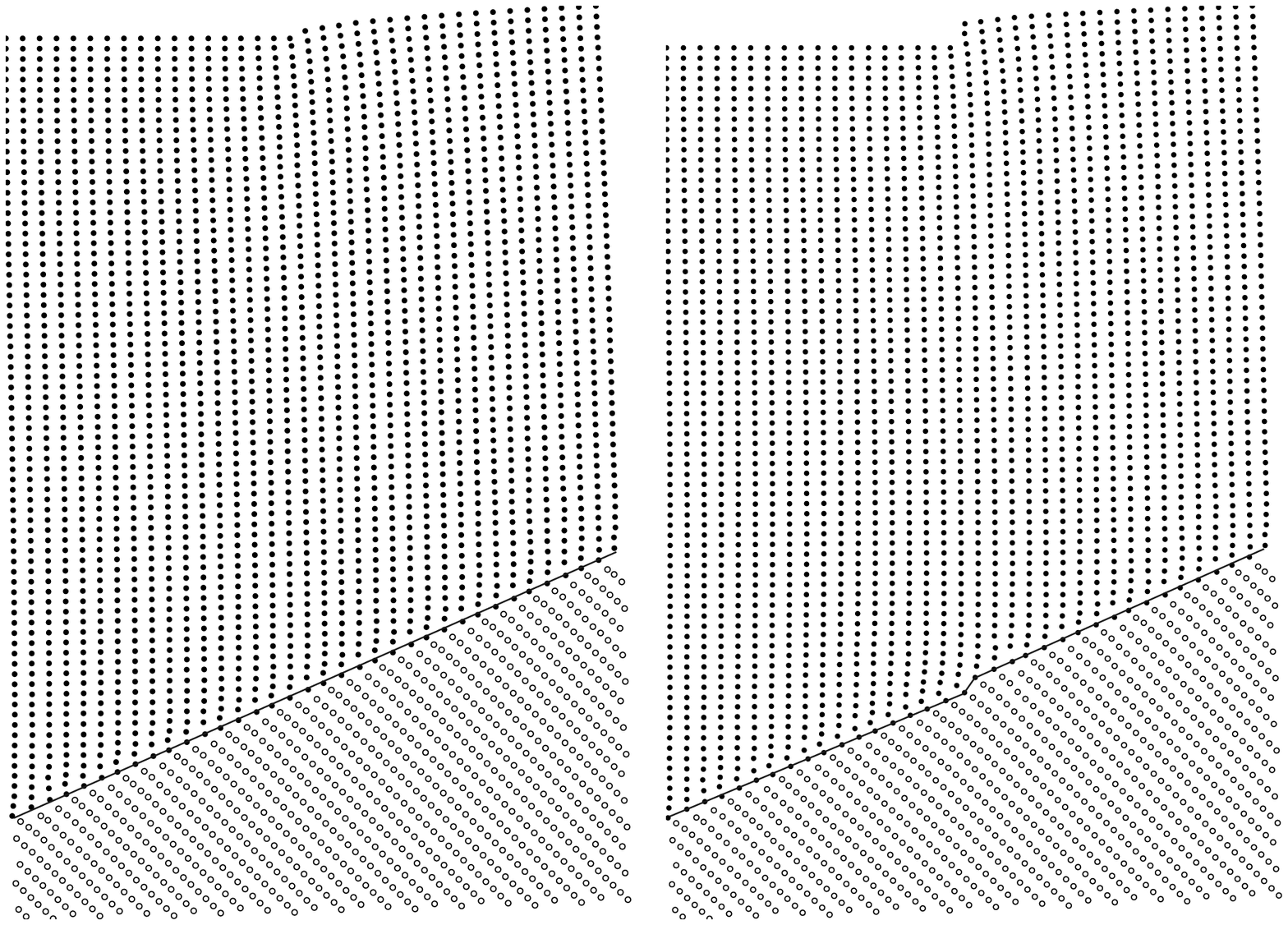}} \hspace{-0.1truein} \epsfysize=3.0truein \epsfbox{\figdir{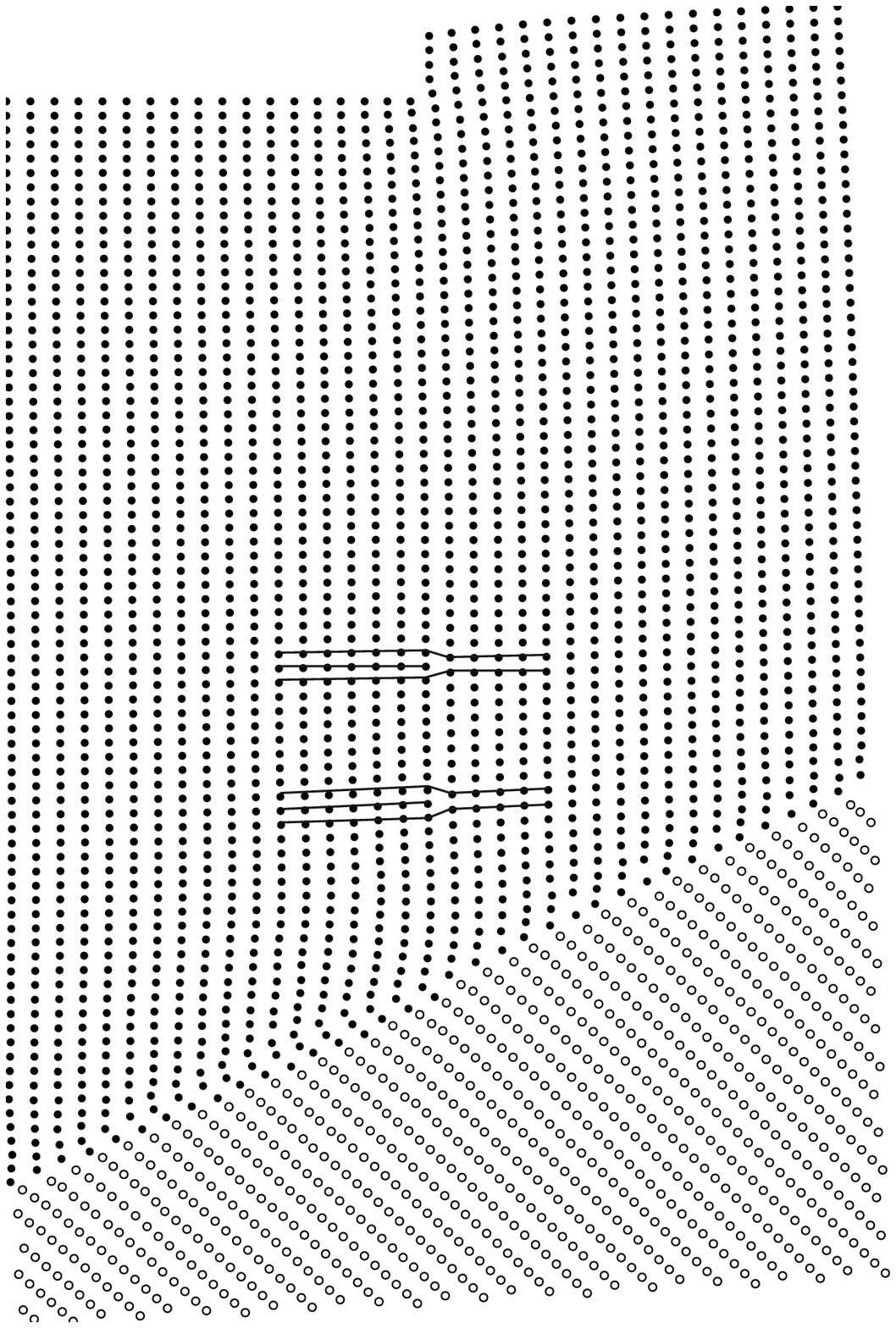}} }    \centerline{(a) \hspace{1.8truein} (b)   \hspace{1.8truein} (c) }
\figcaption{Snapshots of atomic configurations depicting the interaction of dislocations with a grain boundary. (a) Atomic configuration immediately before the nucleation of the partials. (b) Atomic configuration immediately after the nucleation of the first set of partials which have been absorbed into the boundary. (c) The second pair of nucleated partials form a pile up. } \label{gindsnap}
\end{figure}

On application of the load, the bicrystal undergoes some initial
elastic deformation and the first dislocation is nucleated  when
the displacement of the top face reaches 14.2 \AA. 
 This dislocation is driven into the boundary and
is absorbed  without an increase in the load
level. Figs.~\ref{gindsnap}a,b shows the configuration of the grain boundary
immediately before and after this nucleation event. It is seen that
the dislocation absorption produces a step on the grain boundary. This
process may be understood based on the DSC lattice by decomposing the
Burgers vector into DSC lattice vectors (King and Smith, 1980). In our
case, we find that
\bea
\frac{a_o}{2}[\bar1,1,0] = \frac{a_o}{14}[\bar{3} \bar{1} \bar{2}]
+\frac{a_o}{7}[\bar{2} 4 1] 
\eea 
where $\frac{a_o}{14}[\bar{3} \bar{1}\bar{2}]$ is the Burgers vector
of a grain boundary dislocation parallel to the boundary and
$\frac{a_o}{7}[\bar{3}41]$ is perpendicular to the boundary plane. A
careful examination of \fig{gindsnap}b reveals that
$\frac{a_o}{14}[\bar{3} \bar{1}\bar{2}]$ travels along the boundary,
and stops on reaching the end of the nonlinear zone. On subsequent
loading, another pair  of Shockley partials are nucleated when
the displacement of the rigid indenter is 18.2\AA, which again does
not result in any significant reduction of the load.  Unlike the first
pair, these dislocations are not immediately absorbed by the boundary,
and they form a pile-up ahead of the boundary as shown in
\fig{gindsnap}c.  On additional indentation, these dislocations are
also absorbed by the boundary. The simulation was terminated at this
stage.

 The neighboring crystal shows no significant dislocation
activity and thus it may be concluded that slip is not transmitted into
the neighboring grain across the boundary.  The
absorption of the dislocation resulted in a sliding motion of the grain
boundary by the passage of a grain boundary dislocation  and the formation 
of a step on the grain boundary. The formation
of the step appears to result in the increased resistance of the boundary to 
dislocations, as is clear from the fact that a significantly higher stress 
level had to be attained before the absorption of the second dislocation. 

It is worth noting the significant computational
saving obtained by the use of the QC method for this problem. The number of
degrees of freedom used in the QC model was about $10^4$ while a complete atomistic
model of this problem would have required more than $10^7$ degrees of freedom. 
The QC simulation
required  about 140 hours on a DEC-Alpha work-station while a purely atomistic
model would have required a parallel supercomputer.

\section{Conclusions}

This paper was set forth with a few main objectives.  First, the
the quasicontinuum formalism as given
here was advanced as a basis for considering problems involving
multiple grains.  The viewpoint adopted  is that of thinning of
degrees of freedom, with regions far away from defect cores treated
approximately by virtue of finite element interpolation and associated
quadrature rules for evaluating the discrete sums needed to obtain the
total energy.  These ideas also serve as the basis for
the extension of the method to three dimensions and to the
incorporation of dynamic effects via a finite temperature algorithm.

The second main objective of the present paper was to validate the
method in the context of a number of new problems.  In particular, we
have seen that the method allows for the treatment of interfacial
structures and the study of deformation processes that involve interfaces.
Other problems, such as the formation of dislocation junctions and the interactions of cracks and grain boundaries which have also been treated using the ideas presented here will be described in forthcoming papers.

\section{Acknowledgements}
We thank C. Briant, R. Clifton, B. Gerberich, P. Hazzledine, S. Kumar
and A. Schwartzman for many stimulating discussions,  S.W. Sloan for
use of his Delaunay triangulation code and  M.  Daw and S. Foiles
for use of their Dynamo code.  This work was supported by the AFOSR through
grant F49620-95-I-0264,  the NSF through grants CMS-9414648 and
DMR-9632524 and the DOE through grant DE-FG02-95ER14561.  RM
acknowledges support of the NSERC.

\section*{References}
\noindent
\begin{description}
\item{} Ackland, G.~J., Tichy, G., Vitek, V., Finnis, M.~W.~(1987): {\em Phil. Mag. }, {\bf A56}, p.~735.
\item{} Belytschko, T., Tabbara, M.(1993): {\em
Itnl. J. Num. Meth. Engng.}, {\bf 36}, p. 4245.
\item{} Barenblatt, G.~I.~(1962): {\em Adv.~Appl.~Mech.}, {\bf 7}, p.~55.
\item{} Chadwick, P.~(1976): {\em Continuum Mechanics}, John Wiley \&
Sons, New York.
\item{} Christian, J.~W.~(1983): {\em Met. Trans.}, {\bf A14}, p.~1237.
 \item{} Dahmen, U., Hetherington, C.~J.~D., Okeefe, M.~A., Westmacott,
K.~H., Mills, M.~J., Daw, M.~J., Vitek, V.~(1990): {\em Phil.~Mag.~Letters.}, {\bf 62},  p.~327.
\item{} Daw, M.~S.,  Baskes, M.~I.~(1983): {\em Phys.~Rev.~Lett.},
{\bf 50}, p.~1285.
\item{} Ercolessi, F., Adams, J.~(1993): {\em Europhys.~Lett.}, {\bf
26}, p.~583.
\item{}Ericksen, J.~L.~(1984): in {\em Phase Transformations and Material Instabilities in Solids}, edited by M. Gurtin, Academic Press, p.~61.
\item{}Gerberich, W.~W., Nelson, J.~C.,  Lilleodden, E.~T.,  Anderson, P., Wyrobek, J.~T.~(1996):{\em Acta.~Mat.}, {\bf 44}, p.~3585.
\item{}Hirth, J.~P., Lothe, J.~(1968):{\em Theory of Dislocations}, McGraw--Hill, New York.
\item{}Hull, D., Bacon, D.~J.~(1992):{\em Introduction to Dislocations}, Pergamon Press, Oxford.
\item{} King, A.~H., Smith, D.~A.~(1980): {\em Acta.~Cryst.}, {\bf
A36}, p.~335.
\item{} Kohlhoff, S., Gumbsch, P., Fischmeister, H.~F.~(1991): {\em Phil.~Mag.}, {\bf A64}, p.~851.
\item{} Okabe, A., Boots, B., Sugihara, K.~(1992): {\em Spatial Tessellations}, Wiley \& Sons, New York.
\item{} Papadrakakis, M., Ghionis, P.~(1986): {\em Comp.~Meth.~Appl.~Mech.~Engg.}, {\bf 59}, p.~11
\item{} Peierls, R.~E.~(1940): {\em Proc.~Phys.~Soc.~Lond.}, {\bf 52}, p.~34.
\item{} Shenoy, V.~B., Phillips, R.~(1997):{\em Phil.~Mag.}, {\bf A76}, p.~367.
\item{} Sloan, S.~W.~(1993): {\em Computers and Structures}, {\bf 47}, p.~441.
\item{} Tadmor, E., Ortiz, M., Phillips, R.~(1996): {\it Phil.~Mag.}, {\bf A73}, p.~1529. 
\item{} Tadmor, E.~(1996): Ph.D. Thesis, Brown University.
\item{}Tadmor, E.~B., Miller, R., Phillips, R., Ortiz, M.(1997): To be submitted to {\em Acta.~Met.}
\item{} Thomson, R., Zhou, S.~J., Carlsson, A.~E.,  and Tewary,
V.~K.~(1992): {\em Phys.~Rev.~B}, {\bf 46}, p.~10613.
\item{} Xu, X.-P., Argon, A.~S., Ortiz, M.~(1995): {\em
Phil. Mag.}, {\bf A72}, p.~415.
\item{} Zienkiewicz, O.~C., Zhu, J.~Z.~(1987): {\em Itnl.~J.~Num.~Meth.~Engg.}, {\bf 24}, p.~337.
\end{description}

\newpage
\begin{table}[h]
\bc
\bt{lcc}
\hline
Layer  & LS (eV) & QC (eV) \\
\hline
A & 0.3433 &  0.3433  \\
B & 0.0368 &  0.0367  \\
C & 0.0024 &  0.0025  \\
D & 0.0003 &  -0.0003 \\
E & 0.0000 &   0.0000 \\
F & 0.0000 &   0.0000 \\
\hline
\et
\ec
\caption{Comparison of energies of atoms as obtained from LS and QC
for the (111) free surface. Layer $A$ is the outermost layer.}
\label{SurfTable}
\end{table}

\begin{table}[h]
\bc
\bt{lcc}
\hline
Layer  & LS (eV) & QC (eV) \\
\hline
A & -0.0051 &  -0.0051  \\
B & 0.0122 &  0.0123  \\
C & 0.0031 &  0.0031  \\
D & 0.0000 &  0.0000  \\
E & 0.0000 & 0.0000  \\
F & 0.0000 & 0.0000  \\
\hline
\et
\ec
\caption{Comparison of energies of atoms as obtained from LS and QC for
 the twin boundary.}
\label{TwinTable}
\end{table}

\newpage
\listoffigures

\newpage


\end{document}